\documentclass[runningheads]{llncs}

\usepackage[T1]{fontenc}
\usepackage{amsmath}
\usepackage{threeparttable}
\usepackage{xcolor}
\usepackage{graphicx}
\usepackage{tikz}
\usepackage{caption}
\usepackage{subcaption}
\usepackage{booktabs}
\usepackage{xurl}
\usepackage{hyperref}
\usepackage{marvosym}

\begin{document}
\title{PassTSL: Modeling Human-Created Passwords through Two-Stage Learning}

\author{
Haozhang Li\inst{1} \and 
Yangde Wang\inst{1(\textrm{\Letter})}\orcidID{0009-0005-5813-1470}\thanks{The first two co-authors, Y.~Wang and H.~Li, contributed equally to this work.} \and 
Weidong Qiu\inst{1}\orcidID{0000-0001-6428-1655} \and 
Shujun Li\inst{2}\orcidID{0000-0001-5628-7328} \and 
Peng Tang\inst{1}\orcidID{0000-0001-6607-1280}
}
\authorrunning{Y.~Wang and H.~Li et al.}

\institute{School of Cyber Science and Engineering, Shanghai Jiao Tong University, Shanghai, China\\ 
\email{thomas\_li, softds, qiuwd, tangpeng@sjtu.edu.cn} 
\and School of Computing \& Institute of Cyber Security for Society (iCSS), University of Kent, Canterbury, Kent, CT2 7NP, UK\\ 
\email{S.J.Li@kent.ac.uk}}

\maketitle

\begin{abstract}
Textual passwords are still the most widely used user authentication mechanism. Due to the close connections between textual passwords and natural languages, advanced technologies in natural language processing (NLP) and machine learning (ML) could be used to model passwords for different purposes such as studying human password-creation behaviors and developing more advanced password cracking methods for informing better defence mechanisms. In this paper, we propose PassTSL (modeling human-created \underline{Pass}words through \underline{T}wo-\underline{S}tage \underline{L}earning), inspired by the popular pretraining-finetuning framework in NLP and deep learning (DL). We report how different pretraining settings affected PassTSL and proved its effectiveness by applying it to six large leaked password databases. Experimental results showed that it outperforms five state-of-the-art (SOTA) password cracking methods on password guessing by a significant margin ranging from 4.11\% to 64.69\% at the maximum point. Based on PassTSL, we also implemented a password strength meter (PSM), and our experiments showed that it was able to estimate password strength more accurately, causing fewer unsafe errors (overestimating the password strength) than two other SOTA PSMs when they produce the same rate of safe errors (underestimating the password strength): a neural-network based method and zxcvbn. Furthermore, we explored multiple finetuning settings, and our evaluations showed that, even a small amount of additional training data, e.g., only 0.1\% of the pretrained data, can lead to over 3\% improvement in password guessing on average. We also proposed a heuristic approach to selecting finetuning passwords based on JS (Jensen-Shannon) divergence and experimental results validated its usefulness. In summary, our contributions demonstrate the potential and feasibility of applying advanced NLP and ML methods to password modeling and cracking.

\end{abstract}

\section{Introduction}
\label{sec:introduction}

Textual passwords are currently the most common authentication scheme~\cite{web_auth} and will continue to be widely used in the foreseeable future~\cite{agenda}. However, human-created passwords are often vulnerable to attacks based on data-driven probabilistic models~\cite{personal_info,sepcfg,birthday}. Current state-of-the-art (SOTA) modeling approaches include the Markov chain based models (modeling $n$-gram transfer probabilities)~\cite{markov_origin,pwd_strength_markov,adaptivePM_markov,backoff,omen}, pattern-based models (modeling password semantic structures)~\cite{pcfg_origin,pcfg_pinyin,sepcfg,pcfg_keyboard,pcfg_digit,chunk_level}, recurrent neural network (RNN) based models (learning to predict transfer distributions using complete preceding context)~\cite{fla}, and generative adversarial network (GAN) based models (adversarially learning the overall representation of a password set)~\cite{passgan,genpass,dpg}.

For human-created textual passwords have to be memorized by the human users, they are often created partly or even fully based on natural languages so that there are elements reflecting natural language semantics. However, there are also substantial differences between human-created textual passwords and natural languages, e.g., the former do not usually include any white spaces or other obvious separator characters like in natural languages and are much shorter. Despite the differences, many natural language processing (NLP) and machine learning (ML) techniques can still be applied to password modeling and prediction, e.g., the widely used sequence2sequence prediction in NLP can be easily generalized to human-created textual passwords so that the masked part of a password like `q1w2e[MASK]' can be predicted to be more likely `q1w2e3' rather than other characters.

Inspired by the pretraining-finetuning framework that have been widely used for NLP and deep learning (DL) models in recent years, this paper presents our work about PassTSL (modeling human-created \underline{Pass}words through \underline{T}wo-\underline{S}tage \underline{L}earning), a deep learning based password model powered by the self-attention mechanism in transformers~\cite{transformer} under a two-staged learning process: pretraining based on a large and more general database, and finetuning based on a smaller and more specific database for the target password database.

We conducted extensive experiments on the impact of the network size, the training data used, and the training data size on the password distribution modeling ability in the pretraining phase. We implemented PassTSL and applied it to six large leaked password databases. Our performance evaluation results manifested the effectiveness of PassTSL, compared against five SOTA password guessing models, including the 6-gram Markov model~\cite{backoff}, Ma et al.'s backoff Markov model~\cite{backoff}, the latest released implementation of the original PCFG-based password cracking method~\cite{pcfg_origin}, Houshmand et al.'s semantic PCFG~\cite{pcfg_keyboard}, and the RNN-based FLA~\cite{fla}. In our experiments, we utilized the Monte Carlo method~\cite{monte_carlo} to evaluate the performance with a large number of guesses, up to $10^{20}$.

Based on PassTSL, we further designed a lightweight password strength meter (PSM) that can estimate the strength of a password in real time. Experiments proved that our PSM was more accurate than the FLA-based PSM~\cite{fla} and the SOTA PSM zxcvbn~\cite{zxcvbn} for password strength estimation.

To further enlarge knowledge learned in the pretraining stage, we explored several finetuning schemes. We discussed and tested the guidance roles of password database properties for finetuning, and showed the distinctive advantage of our two-stage model: the finetuning step was able to improve the performance of password cracking on the target password database using only $10^6$ additional passwords (0.1\% of the database used in the pretraining phase), with a significant margin of up to 3\%. Additionally, we proposed an approach to selecting the finetuning password database based on JS (Jensen-Shannon) divergence~\cite{JSD} between pretraining/candidate finetuning databases and the target database, and validated its effectiveness.

To summarize, the main contributions of this paper are as follows.
\begin{itemize}
\item We propose PassTSL, a neural network model that introduced the self-attention mechanism to promote password modeling performance. PassTSL was able to outperform SOTA password methods in six large databases by a significant margin ranging from 4.11\% to 64.69\% at the maximum point.

\item We introduce the pretraining-finetuning framework into the field of password cracking, for the first time in the literature (to the best of our knowledge), and our experimental results demonstrate its effectiveness.

\item We investigated the impact of two properties of password databases (language and service type) on the performance of the finetuning stage, explored the effectiveness of few-shot learning, and obtained an insight on selecting appropriate finetuning passwords.
\end{itemize}

The rest of the paper is organized as follows. The next section provides an overview of related work. Section~\ref{section:pretraining} introduces PassTSL by reporting the model structure, the impact of pretraining settings, and the performance in comparison to other password cracking methods and PSMs. Section~\ref{section:finetuning} showcases our investigation on finetuning PassTSL, presenting the performance results under various finetuning schemes. The last section concludes our work.

\section{Related Work}

\subsection{Pretrained Models}

To overcome the limitation about the lack of available large-scale datasets, machine learning researchers proposed transfer learning~\cite{transfer_survey} and a two-staged learning framework including the pretraining and finetuning stages~\cite{plmSurvey}. When used for NLP problems, the two-staged framework first encodes linguistic knowledge from a large-scale corpus and then transfers the captured knowledge about the underlying language to a more specified task, which avoids training from scratch but uses a smaller more targeted corpus to finetune the pretrained model~\cite{gpt3}. NLP researchers have proposed various pretrained language models (PLMs) using large unlabeled corpora to learn contextual word embeddings~\cite{plmSurvey}. In 2017 Vaswani et al.~\cite{transformer} proposed Transformer to capture higher-level concepts in context like polysemous disambiguation and syntactic structures. Radford et al.~\cite{gpt} proposed GPT in 2018 and Devlin et al.~\cite{bert} proposed BERT in 2019 for NLP tasks including text sentiment classification, named entity recognition, and Q\&A. Researchers subsequently introduced additional improvements to GPT and BERT, designing new PLMs with better performance such as GPT2~\cite{gpt2}. Some researchers also changed the model architecture and explored new pretraining tasks, leading to work such as BART~\cite{bart} and XLNet~\cite{xlnet}.

\subsection{Password Modeling Methods}

Password guessing attacks are probably as old as the history of passwords. Much research has been done in this area, and we would only discuss methods and contributions that are most relevant to our work.

\textbf{Markov.} Narayanan et al.~\cite{markov_origin} proposed to guess passwords using an $n$-gram Markov model. The model was further extended by Ma et al.~\cite{backoff} and Dürmuth et al.~\cite{omen}. Ma et al.~\cite{backoff} explored normalization techniques and smoothing strategies for Markov models, proposing Laplace smoothing and the backoff mechanism to prevent overfitting of higher-order Markov models. Dürmuth et al.~\cite{omen} optimized the enumeration process by proposing an ordered Markov enumerator (OMEN) based on approximate sorting. However, all Markov model based methods are commonly limited by memory resources. The hyperparameter $n$ is usually taken as 5 or 6 and thus longer-distance contextual information cannot be considered.

\textbf{PCFG.} Weir et al.~\cite{pcfg_origin} explored the use of the probabilistic context-free grammar (PCFG) for password analyses and cracking. They considered passwords as instances of templates based on character types with different terminals (i.e., strings of the same type). For example, the password `alice123!@' is an instance of the template $L_5D_3S_2$, and its probability is calculated as $P(\text{alice123!@})=P(L_5D_3S_2)\times P(\text{alice}|L_5) \times P(\text{123}|D_3) \times P(\text{!@}|S_2)$. Li et al.~\cite{pcfg_pinyin} introduced Pinyin as a new type of password segments to improve Chinese password guessing. Houshmand et al.~\cite{pcfg_keyboard} introduced keyboard patterns and multi-word detection into PCFG to enhance its power. Veras et al.~\cite{sepcfg} used NLP methods to extract semantic information in passwords, and proposed a semantic PCFG that incorporates more types of password segments with different semantic meanings.

\textbf{Neural network based methods.} Melicher et al.~\cite{fla} proposed to model passwords using a long short-term memory (LSTM) network, which is referred by other researchers as FLA (derived from three words in their paper's title: `Fast', `Lean', and `Accurate'). They also designed a PSM based on a highly compressed FLA network. Hitaj et al.~\cite{passgan} proposed PassGAN, which guesses passwords via a GAN. However, they reported that PassGAN required more guesses in order to achieve the same cracking performance as FLA. Dario et al.~\cite{dpg} demonstrated the potential of representation learning in improving PassGAN. In particular, they proposed to dynamically control the latent space of PassGAN through the feedback from correct guesses during password guessing to mimic the unknown distribution of the target passwords.

More recently, some researchers working on password cracking have started recognizing the potential of transformer-based deep learning models. He et al.~\cite{PassTrans} investigated the password reuse problem and presented PassTrans, which was able to guess variants of a given password for online attacks. Similarly, Xu et al.~\cite{Passbert} proposed PassBERT for conditional password guessing, cracking a password fully under the condition of having partial knowledge of the password. However, neither PassTrans nor PassBERT directly models the probability distribution of the target password(s). Rando et al.~\cite{Passgpt} introduced the concept of guided password generation to guess passwords that match arbitrary constraints, but did not consider a finetuning stage.

\section{PassTSL: Pretraining}
\label{section:pretraining}

We first report the architecture and pretraining method of PassTSL. Then, we discuss the effects of different pretraining settings on PassTSL's performance. Finally, we demonstrate advantages of the pretrained (without finetuning) PassTSL over SOTA methods via Monte Carlo simulation from two aspects, password cracking and password strength estimation.

\subsection{Method}
\label{section:pretraining-method}

\begin{figure}[htb]
\centering
\includegraphics[width=0.5\linewidth]{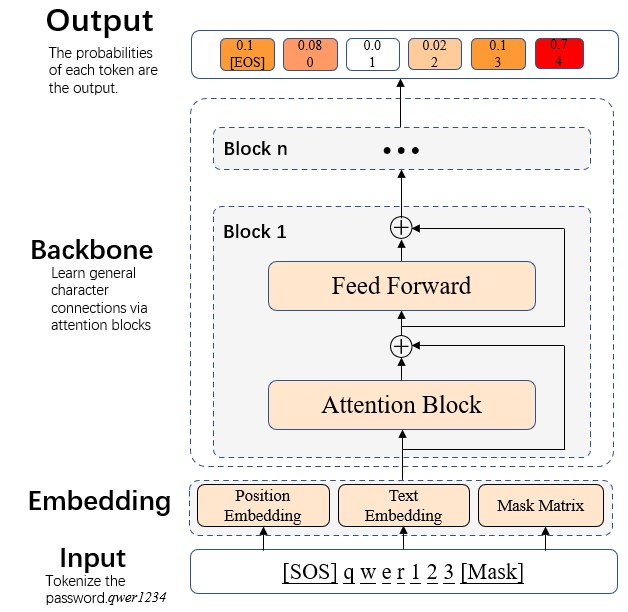}
\caption{The overall architecture of PassTSL. The redder a cell's color is, the more likely it would be the next character. Each module is further explained in Section~\ref{section:pretraining-method}.}
\label{fig:PassTSL}
\end{figure}

Here we propose PassTSL's pretrained model, a base line model of our two-staged password guessing method, in order to learn universal representations of passwords. Formally, password modeling is constructed as an unsupervised distribution estimation of passwords $(\mathbf{x}=\lbrace c_1, c_2, \ldots, c_m\rbrace)$, where $\mathbf{x}$ stands for a password and $\mathbf{c}_i$ is the $i$-th character in $\mathbf{x}$. We use the standard language model object to maximize a likelihood function $L(\mathbf{x})$:
\begin{equation}
L(\mathbf{x}) = \sum_{i=1}^{m} log P(c_i|c_1, \ldots, c_{i-1};\theta),
\label{equ:loss}
\end{equation}
where $\theta$ represents parameters of PassTSL. An overview of the PassTSL architecture is illustrated in Figure~\ref{fig:PassTSL} with an example.

\textbf{Vocabulary and tokenization.} We focus on character-level tokens instead of commonly used word-level tokens in NLP or pattern-level tokens in PCFG. Texts (especially Western texts) are naturally separated by spaces, while passwords rarely include any white spaces or other uniformly defined separators so characters are what we have to start with to analyse passwords.

\textbf{Input.} The input is a single password since PassTSL aims to learn character-level transfer probability distributions. Each password $\mathbf{x}$ will be preprocessed to a sequence of characters after character-level tokenization. A special start-of-sequence ([SOS]) token is added to the beginning of $\mathbf{x}$ as the initial input when generating a candidate or calculating the probability of a given password. A special end-of-sequence ([EOS]) token is also added as the end symbol to help PassTSL learn when to stop decoding.

\textbf{Embedding}. The tokenized sequence will be encoded by the text embedding layer and the position embedding layer. The resulting vectors are then summed up to get the representation of $\mathbf{x}$, and to construct the self-attention mask matrix $\mathbf{M}$ for each password, with element $\mathbf{M}_{ij}$ satisfying
\begin{equation}
\begin{aligned}
\mathbf{M}_{ij} = 
\begin{cases}
0 & i \geq j\text{, need attention},\\
-\infty & i < j\text{, need to be masked}.
\end{cases}
\end{aligned}
\label{equ:masked}
\end{equation}

\textbf{Backbone network.} Our model is a multi-layered network based on the architecture of a transformer decoder~\cite{transformer}. The embedding vector $\mathbf{R}_0$ of sequence $\mathbf{x}$ is encoded as a contextual representation $\mathbf{R_L}$ by the $L$-block transformer decoder after normalization. $\mathbf{R_L}$ is fed into a linear classifier to compute the distribution over target tokens. Formally, in the $i$th block ($0 < i \leq L$), a self-attention operation is implemented as:
\begin{equation}
\begin{aligned}
\mathbf{Q}=\mathbf{R}_{i-1}\mathbf{W}_i^Q, \mathbf{K}=\mathbf{R}_{i-1}\mathbf{W}_i^K,  \mathbf{V}=\mathbf{R}_{i-1}\mathbf{W}_i^V,
\end{aligned}
\label{equ:attention1}
\end{equation}
\begin{equation}
\begin{aligned}
Attention(\mathbf{Q}, \mathbf{K}, \mathbf{V}) = softmax(\frac{\mathbf{Q}\mathbf{K}^T}{\sqrt{d_k}}+\mathbf{M})\mathbf{V},
\end{aligned}
\label{equ:attention2}
\end{equation}
where $\mathbf{R}_{i-1}$ is the output of the $i-1$-th block, while $\mathbf{W}_i^Q$, $\mathbf{W}_i^K$, $\mathbf{W}_i^V$ are the parameter matrices for linearly mapping $\mathbf{R}_{i-1}$ to a triple. $\mathbf{M}$ is the self-attention mask matrix fed into PassTSL together with the context tokens.

\textbf{Hyperparameters.} PassTSL is defined by the following hyperparameters.

\begin{itemize}
\item[*] $l$: \textit{Number of layers}, which represents the number of decoder blocks.

\item[*] $E$: \textit{Embedding size}, which determines the number of dimensions for embedding layers and the hidden state in each block.

\item[*] $I$: \textit{Intermediate size}, which indicates dimensions of the feed-forward layer in each block.

\item[*] $h$: \textit{Number of heads}, which represents the number of self-attention heads.

\item[*] \textit{Vocab size}, which represents the size of the PassTSL vocabulary. The vocabulary is composed of 95 printable ASCII characters and five special characters ([PAD], [SOS], [EOS], [UNK], and [MASK]).

\item[*] \textit{Attention dropout}, which specifies the dropout ratio of the attention blocks. It is instantiated to 0.1.
\end{itemize}

\textbf{Model size.} To balance the computational costs with the performance of PassTSL, we design two PassTSL instances of different sizes, shown in Table~\ref{tab:model_size}. We will report their performance in Section~\ref{section:pretraining-settings}.

\begin{table}[htb]
\centering
\caption{Structure configurations of different PassTSL instances.}
\label{tab:model_size}
\begin{threeparttable}
    \begin{tabular}{c c c c c c}
        \toprule
        Model & $l$ & $E$ & $I$ & $h$ & \#(All Parameters)\\
        \midrule
        $\text{PassTSL}_{\text{Base}}$ & 12 & 768 & 3,072 & 12 & 85,919,232\\
        $\text{PassTSL}_{\text{Small}}$ & 6 & 256 & 1,024 & 4 & 4,781,056\\
        \bottomrule
    \end{tabular}
\end{threeparttable} 
\end{table}

\textbf{Password generation.} Evaluating performances on the target password databases by enumeration is inefficient and too resource-intensive. For instance, 100 million ($10^8$) guessed passwords would occupy around 1GB memory. Besides, some password cracking models are particularly insufficient in this regards, e.g., Pasquini et al.~\cite{dpg} pointed out that it would take more than two weeks to generate $10^{10}$ passwords using FLA~\cite{fla}. Therefore we decided to use Monte Carlo simulation to estimate the number of guesses required for a given target password.

\begin{table}[htb]
\centering
\caption{Summary of password datasets used.}
\label{tab:pre_data}
\small
\begin{threeparttable}
        \begin{tabular}{c c c c c c c}
        \toprule
        Name & Service & User Type & Year & \#(Passwords) & Length$\leq$5 & Usage\\
        \midrule
        COMB & Multiple & Mixed & 2021 & 3.3B & 4.1\% & Pretrain\\
        \midrule
        CSDN & Social Forum & CN (Chinese) & 2011 & 6.4M & 0\% & Pretrain\\
        17173 & Entertainment & CN & 2011 & 17.9M & 0\% & Target\\
        178 & Entertainment & CN & 2011 & 9.1M & 0\% & Target\\
        Tianya & Social Forum & CN & 2011 & 30.3M & 0.0001\% & Target\\
        \midrule
        Gmail & Life & EN (English) & 2014 & 4.7M & 0.004\% & Pretrain\\
        MyHeritage & Life & EN & 2018 & 84.8M & 0.02\% & Target\\
        RockYou & Social Forum & EN & 2009 & 28.7M & 0.0008\% & Target\\
        Twitter & Social Forum & EN & 2016 & 67.1M & 0\% & Target\\
        \bottomrule
    \end{tabular}
\end{threeparttable}
\end{table}

\begin{table}[!htb]
\centering
\caption{Settings for pretraining PassTSL instances.}
\label{tab:pretrain_settings}
\begin{threeparttable}
    \begin{tabular}{c c c c c}
        \toprule
        Name & Model Size$^a$ & Data & Data Size & Target Passwords\\
        \midrule
        $\text{PassTSL}_{\text{Small}}^{\text{COMB\_100M}}$ & Small & COMB & 100M & All passwords\\
       $\text{PassTSL}_{\text{Small}}^{\text{COMB\_1M}}$ & Small & COMB & 1M & All passwords\\
        $\text{PassTSL}_{\text{Base}}^{\text{CSDN\_1M}}$ & Base & CSDN & 1M & Chinese Passwords\\
        $\text{PassTSL}_{\text{Small}}^{\text{CSDN\_1M}}$ & Small & CSDN & 1M & Chinese Passwords\\
        $\text{PassTSL}_{\text{Base}}^{\text{Gmail\_1M}}$ & Base & Gmail & 1M & English Passwords\\
        $\text{PassTSL}_{\text{Small}}^{\text{Gmail\_1M}}$ & Small & Gmail & 1M & English Passwords\\
        \bottomrule
    \end{tabular}
    \begin{tablenotes}
    \footnotesize
    \item[a] See Table~\ref{tab:model_size} for hyperparameters of the \textit{base} and \textit{small} models.
    \end{tablenotes}
\end{threeparttable}
\end{table}

\textbf{Datasets.} We employed multiple leaked password databases, four password databases leaked from Chinese websites, and another four leaked from English websites\footnote{The four websites are all run by US-based companies and their users are from many countries and speak many different languages. However, English is usually the dominating or common language used by all users.}, for pretraining and finetuning PassTSL, and also for evaluating its performance against other SOTA methods. These eight databases cover three types of online services and two mainly used languages. Another hybrid database COMB is a compilation of existing data which contains approximately 3.2 billion unique passwords from multiple previous leaks and breaches\footnote{\url{https://cybernews.com/news/largest-compilation-of-emails-and-passwords-leaked-free}}. Table~\ref{tab:pre_data} provides a detailed description.

Although some databases were leaked in 2009 and 2011, we still consider them representative based on some research looking at human users' password creation behaviors, e.g., Bonneau~\cite{pwd_no_change} reported that such behaviors had changed only slightly between 1990 and 2011, and other more recent research~\cite{meter_acc,guidance_pass_compliance} revealed that password policies and practices implemented on top sites had rarely changed.

The databases we used have been widely used by password researchers~\cite{backoff,fla,birthday,chunk_level,pcfg_pinyin}. Since passwords can include sensitive personal information, in this paper we only report aggregated results so no personal information will be leaked.

\subsection{Effect of Different pretraining Settings}
\label{section:pretraining-settings}

\subsubsection{Experimental settings}
\label{section:pretraining-settings-experimental}

We randomly selected 100 million passwords and one million passwords from COMB, one million passwords from CSDN, and one million passwords from Gmail as the pretraining data, denoted by COMB\_100M, COMB\_1M, CSDN\_1M, and Gmail\_1M. These passwords consist of only 95 ASCII printable characters and contain no less than 6 characters.

Table~\ref{tab:pretrain_settings} gives detailed settings. Models were pretrained for 10 epochs with a batch size of 256. In Monte Carlo simulations, one million passwords were sampled from each pretrained model to provide valid and accurate estimates. We also modeled up to $10^{20}$ guesses to ensure completeness. It is an overestimation considering that it would take over six months in the real-world generation.

\newcommand\drawlegend[1]{\protect\tikz\protect\draw [color=#1] plot[] coordinates {(0,0) (0.3,0.1) (0.6,0)};}

\begin{figure}[tb]
\centering
\begin{subfigure}[b]{0.3\linewidth}
    \centering
    \includegraphics[width=\linewidth]{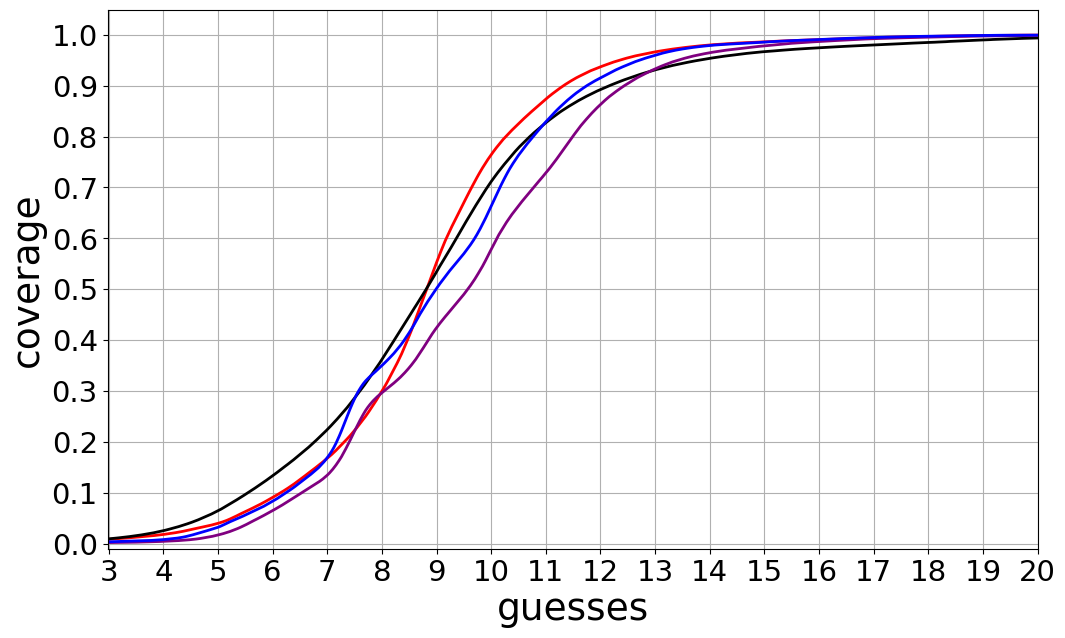}
    \caption{CSDN-17173}
\end{subfigure}
\hfill
\begin{subfigure}[b]{0.3\linewidth}
    \centering
    \includegraphics[width=\linewidth]{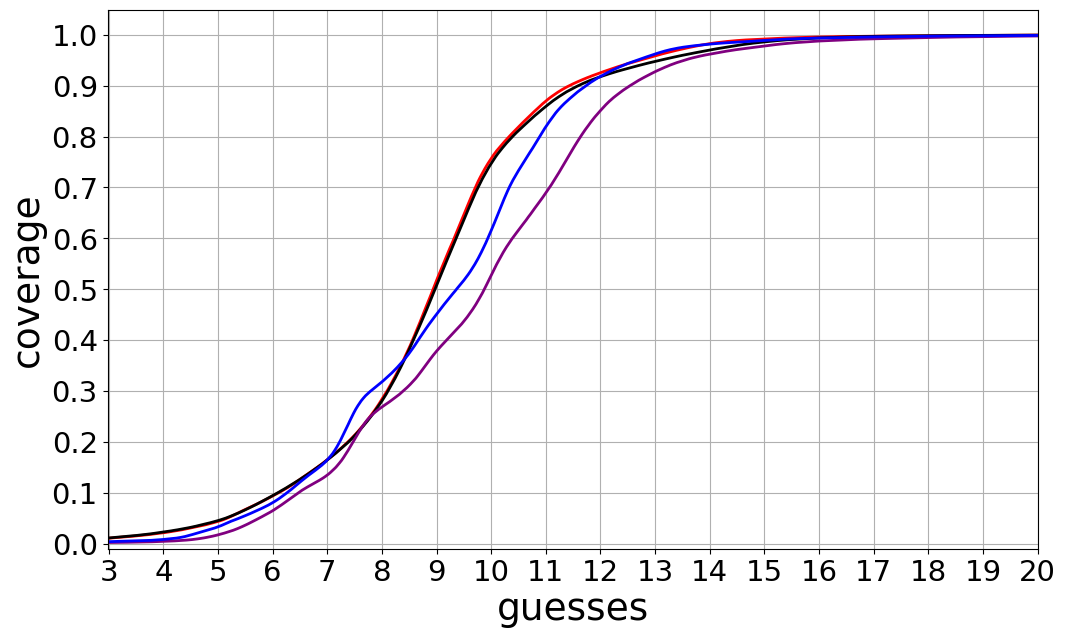}
    \caption{CSDN-178}
\end{subfigure}
\hfill
\begin{subfigure}[b]{0.3\linewidth}
    \centering
    \includegraphics[width=\linewidth]{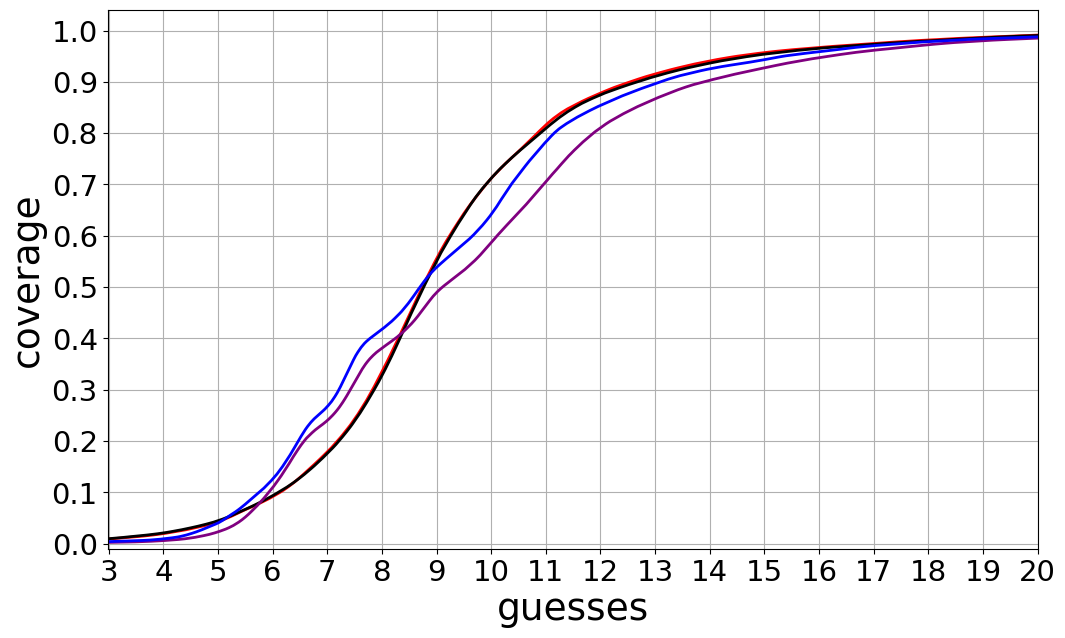}
    \caption{CSDN-Tianya}
\end{subfigure}
\caption{Various pretraining settings for PassTSL attacking Chinese databases: \drawlegend{red} $\text{PassTSL}_{\text{Small}}^{\text{CSDN\_1M}}$, \drawlegend{black} $\text{PassTSL}_{\text{Base}}^{\text{CSDN\_1M}}$, \drawlegend{violet}$\text{PassTSL}_{\text{Small}}^{\text{COMB\_1M}}$, \drawlegend{blue} $\text{PassTSL}_{\text{Small}}^{\text{COMB\_100M}}$. The x-axes represent the number of guesses in the log scale. The y-axes show the percentage of correctly guessed passwords.}
\label{fig:pretrain-1}
\end{figure}

\begin{figure*}
\centering
\begin{subfigure}[b]{0.3\linewidth}
    \centering
    \includegraphics[width=\linewidth]{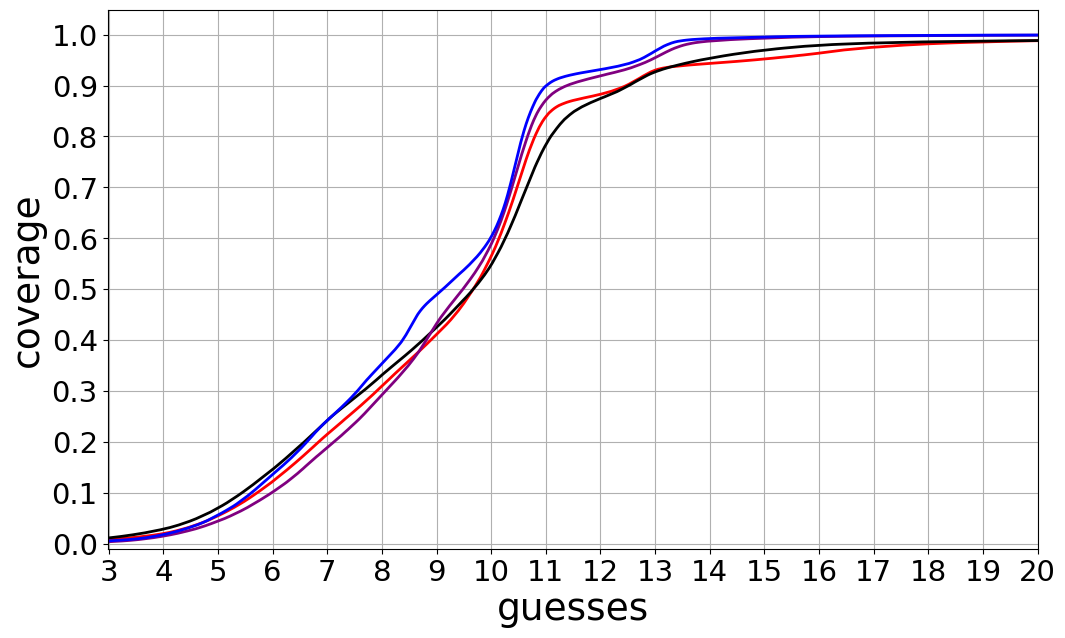}
    \caption{Gmail-MyHeritage}
\end{subfigure}
\hfill
\begin{subfigure}[b]{0.3\linewidth}
    \centering
    \includegraphics[width=\linewidth]{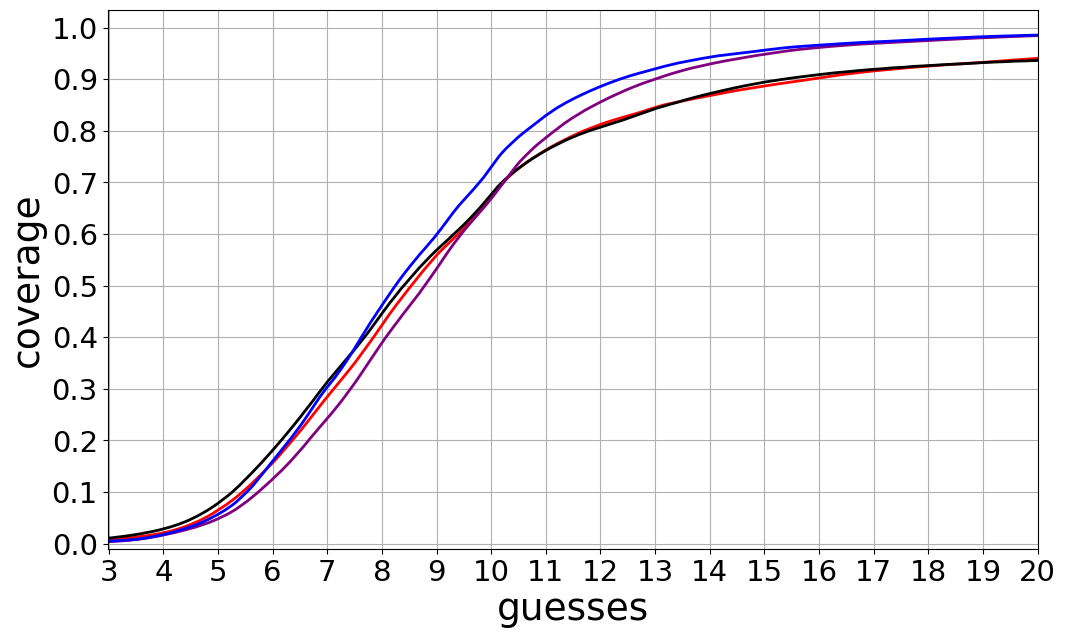}
    \caption{Gmail-RockYou}
\end{subfigure}
\hfill
\begin{subfigure}[b]{0.3\linewidth}
    \centering
    \includegraphics[width=\linewidth]{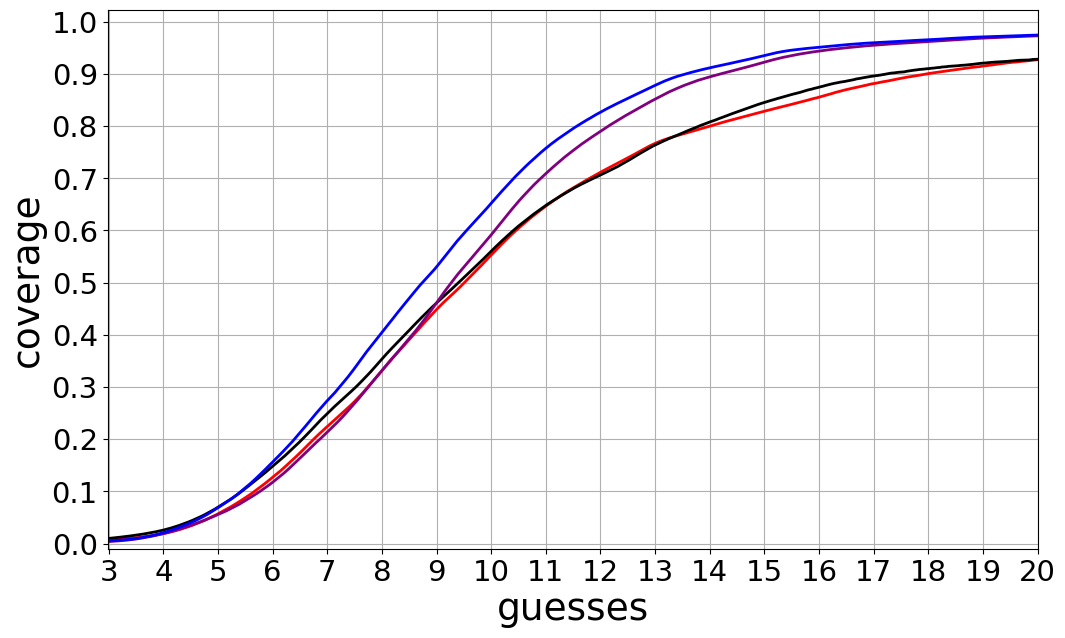}
    \caption{Gmail-Twitter}
\end{subfigure}
\caption{Various pretraining settings for PassTSL attacking English databases: \drawlegend{red} $\text{PassTSL}_{\text{Small}}^{\text{Gmail\_1M}}$, \drawlegend{black} $\text{PassTSL}_{\text{Base}}^{\text{Gmail\_1M}}$, \drawlegend{violet}$\text{PassTSL}_{\text{Small}}^{\text{COMB\_1M}}$, \drawlegend{blue} $\text{PassTSL}_{\text{Small}}^{\text{COMB\_100M}}$.}
\label{fig:pretrain-1-more}
\end{figure*}

\subsubsection{Analyses and Discussions}
\label{section:pretraining-settings-2}

Figures~\ref{fig:pretrain-1} and \ref{fig:pretrain-1-more} show the performance of PassTSL on six testing databases under different pretraining settings. We have the following findings and conclusions.

\textbf{The increase of the model size plays a positive but has a limited effect on the guessing performance.} As shown in Figure~\ref{fig:pretrain-1}, the curves of $\text{PassTSL}_{\text{Base}}^{\text{CSDN\_1M}}$ and $\text{PassTSL}_{\text{Small}}^{\text{CSDN\_1M}}$ largely overlap. Figure~\ref{fig:pretrain-1-more} indicates that the simulation curves of $\text{PassTSL}_{\text{Base}}^{\text{Gmail\_1M}}$ are a bit higher than those of $\text{PassTSL}_{\text{Small}}^{\text{Gmail\_1M}}$, specifically \textbf{0.92\%} on average at $10^{20}$ guesses. However, such improvement is not surprising considering that the number of parameters for $\text{PassTSL}_{\text{Base}}$ is 20 times larger than that for $\text{PassTSL}_{\text{Small}}$. For sequences like passwords with shorter length and less logical information than texts in general NLP tasks, a small-scale model has been already strong enough to sufficiently learn the inner linguistic features. Larger-scale models, while still likely to improve the effectiveness of password modeling, have to bear the risk of overfitting.

\textbf{Pretraining PassTSL with mixed-language passwords is able to help better modeling English passwords.} Figure~\ref{fig:pretrain-1-more} shows that mixed-language based PassTSL instances obviously outperformed those pretrained using English passwords only. Particularly, the cracking rate of $\text{PassTSL}_{\text{Small}}^{\text{COMB\_1M}}$ is on average \textbf{3.38\%} higher than that of $\text{PassTSL}_{\text{Base}}^{\text{Gmail\_1M}}$ and \textbf{3.46\%} higher than that of $\text{PassTSL}_{\text{Small}}^{\text{Gmail\_1M}}$, while $\text{PassTSL}_{\text{Small}}^{\text{COMB\_100M}}$ is on average \textbf{3.48\%} and \textbf{3.55\%} higher than the same two benchmarks, respectively. This significant performance gain does not exist when such PassTSL instances are used to attack Chinese databases, as shown in Figure~\ref{fig:pretrain-1}. We believe that it is because hybrid databases like COMB are more dominated by English passwords, making the pretrained model more biased towards English passwords. Such a bias also implies that using Chinese passwords for finetuning can potentially improve a COMB-pretrained PassTSL model's performance against Chinese databases, which was proved in our experimental results reported in Section~\ref{section:finetuning-2}.

\textbf{The increase of the training data size will significantly improve the performance.} $\text{PassTSL}_{\text{Small}}^{\text{COMB\_100M}}$ has noticeable improvement over $\text{PassTSL}_{\text{Small}}^{\text{COMB\_1M}}$ on each test set as shown in Figures~\ref{fig:pretrain-1} and \ref{fig:pretrain-1-more}. Besides, although the mixed database COMB is more biased towards English passwords, the performance of $\text{PassTSL}_{\text{Small}}^{\text{COMB\_100M}}$ on Chinese testing sets is still better, likely because the amount of training data from COMB is much larger than the amount of CSDN. As shown in Figure~\ref{fig:pretrain-1}, the simulated cracking rates of $\text{PassTSL}_{\text{Small}}^{\text{COMB\_100M}}$ compared to $\text{PassTSL}_{\text{Base}}^{\text{COMB\_100M}}$ differ by only \textbf{0.004\%} (17173), \textbf{0.06\%} (178), and \textbf{0.35\%} (Tianya) at $10^{20}$, respectively. These results suggest that the data-driven PassTSL model possesses the potential to further improve the accuracy of password modeling given more training resources.

Considering the resources required for pretraining, password generation and probabilities calculation, we believe that $\text{PassTSL}_{\text{Small}}^{\text{COMB\_100M}}$ is a good balanced representation of PassTSL's ability to model passwords.

\subsection{Evaluations on Password Cracking and Password Strength Estimation}
\label{section:pretraining-3}

In this subsection, we compare $\text{PassTSL}_{\text{Small}}^{\text{COMB\_100M}}$ with other SOTA password guessing models and PSMs on the testing sets shown in Table~\ref{tab:pre_data}. We briefly describe the models for comparison, and then report their performance.

\begin{figure*}[thb]
\centering
\begin{subfigure}[b]{0.3\linewidth}
    \centering
    \includegraphics[width=\linewidth]{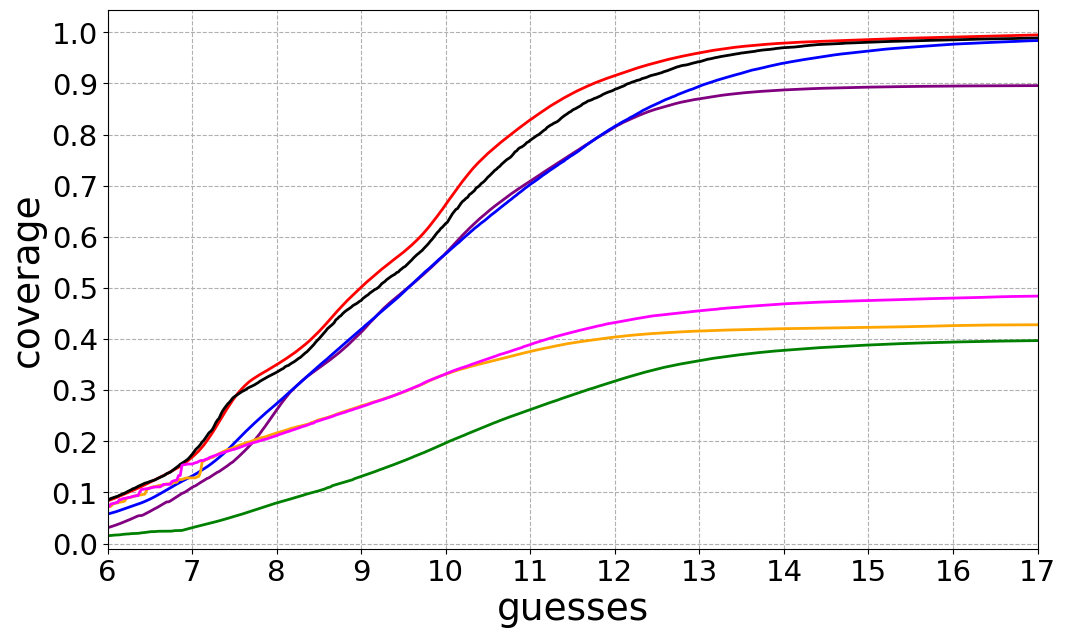}
    \caption{COMB-17173}
\end{subfigure}
\hfill
\begin{subfigure}[b]{0.3\linewidth}
    \centering
    \includegraphics[width=\linewidth]{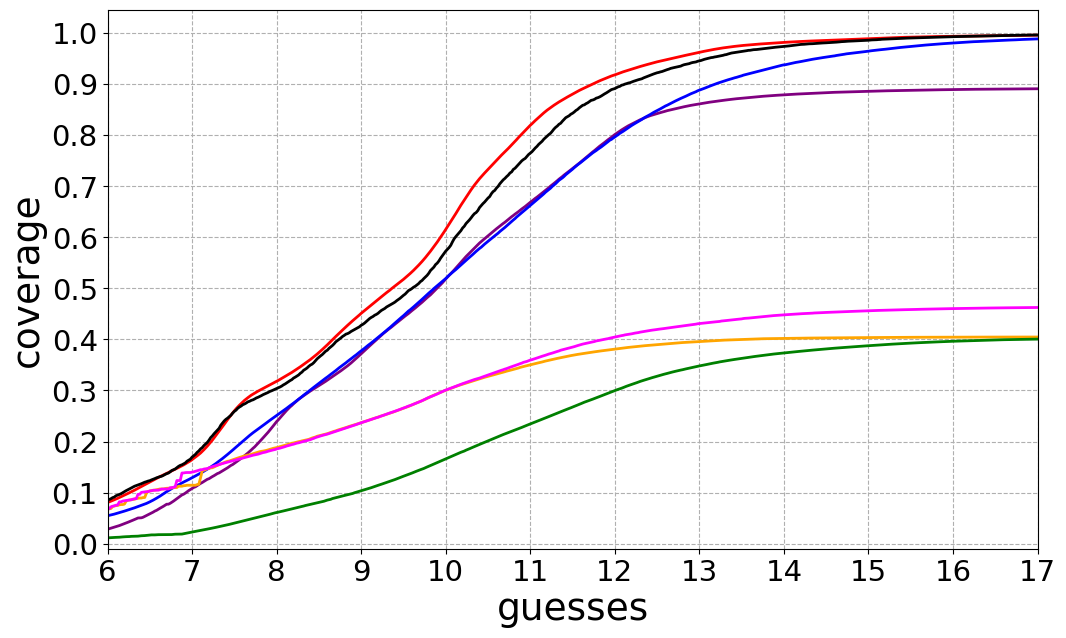}
    \caption{COMB-178}
\end{subfigure}
\hfill
\begin{subfigure}[b]{0.3\linewidth}
    \centering
    \includegraphics[width=\linewidth]{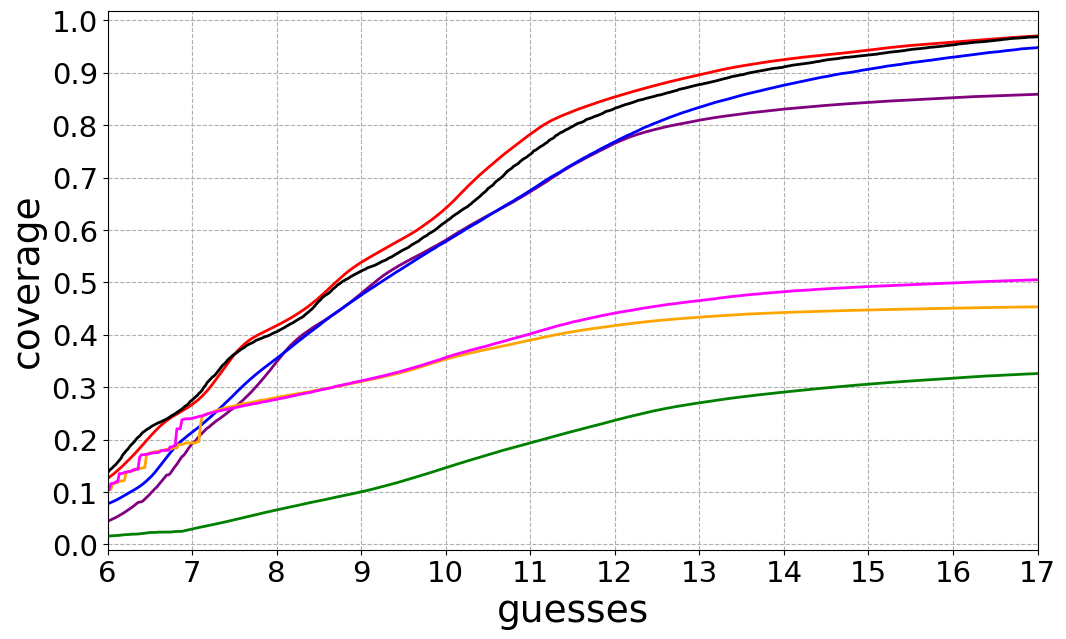}
    \caption{COMB-Tianya}
\end{subfigure}
\\
\begin{subfigure}[b]{0.3\linewidth}
    \centering
    \includegraphics[width=\linewidth]{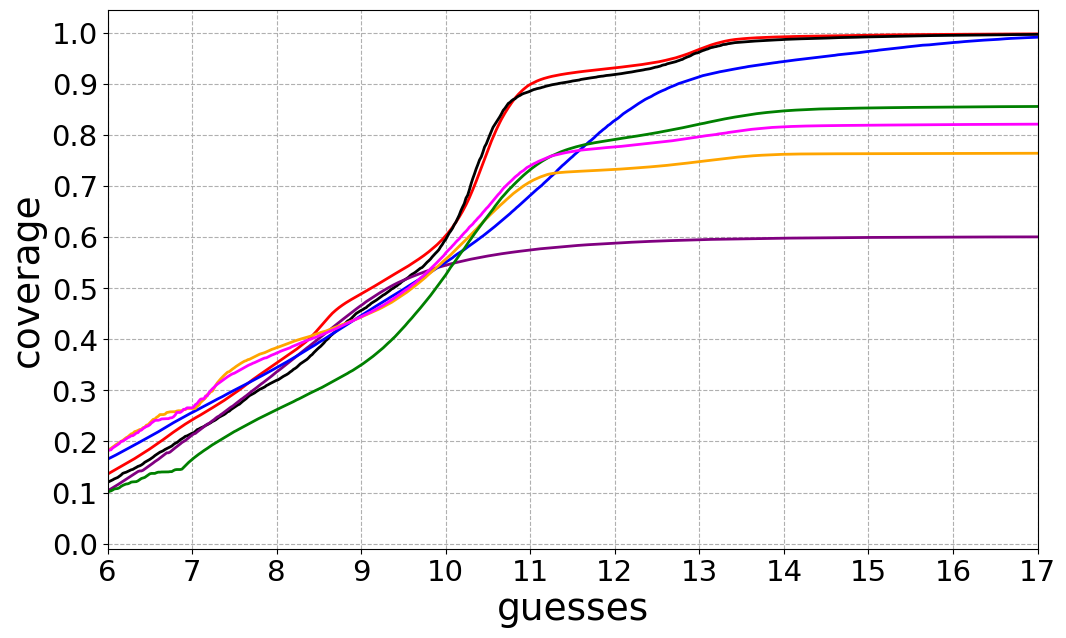}
    \caption{COMB-MyHeritage}
\end{subfigure}
\hfill
\begin{subfigure}[b]{0.3\linewidth}
    \centering
    \includegraphics[width=\linewidth]{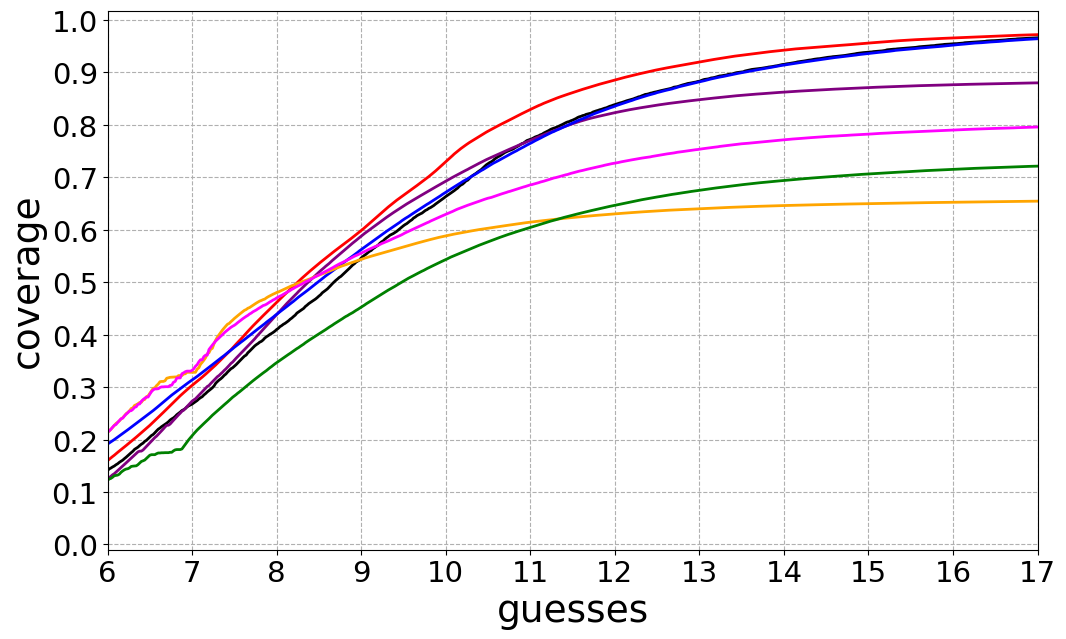}
    \caption{COMB-RockYou}
\end{subfigure}
\hfill
\begin{subfigure}[b]{0.3\linewidth}
    \centering
    \includegraphics[width=\linewidth]{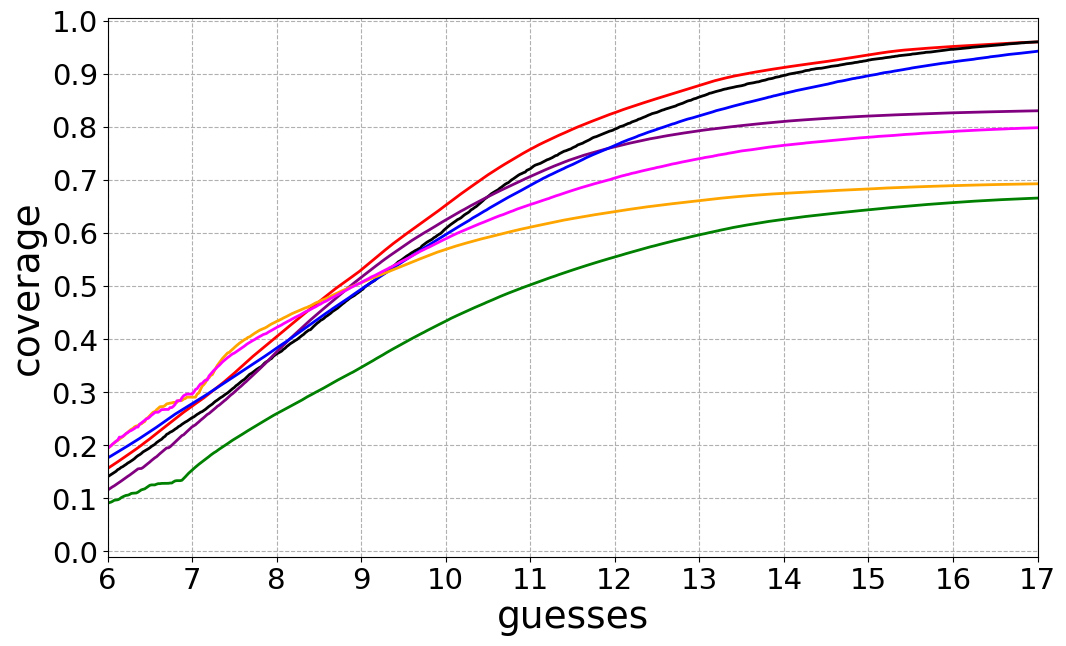}
    \caption{COMB-Twitter}
\end{subfigure}
\caption{Guessing performance of PassTSL against other SOTA models: \drawlegend{red} PassTSL($\text{PassTSL}_{\text{Small}}^{\text{COMB\_100M}}$), \drawlegend{black} FLA, \drawlegend{violet} 6-gram, \drawlegend{blue} Backoff, \drawlegend{orange} PCFG$_{\text{Se}}$, \drawlegend{green} PCFG$_{\text{Ori}}$. The x-axes represent guessing numbers in the log scale. We show the performance under a range of guessed passwords used more commonly by other researchers~\cite{passgan,backoff,Passgpt,fla,sepcfg}.}
\label{fig:pretrain-2}
\end{figure*}

\subsubsection{Password Cracking Tools for Comparison}
\label{section:pretraining-3-1}

The password guessing models compared with PassTSL and their settings are as follows: 1) the 6-gram Markov model used in~\cite{backoff} as a benchmark; 2) the backoff Markov model proposed in~\cite{backoff}; 3) PCFG$_{\text{Se}}$ -- Veras et al.'s semantic PCFG~\cite{sepcfg}; 4) PCFG$_{\text{Ori}}$ -- the latest released (v4.0-rc3) of the original PCFG~\cite{latestpcfg} developed and maintained by Matt Weir, the lead author of the original PCFG paper~\cite{pcfg_origin}; and 5) FLA~\cite{fla} with the default configurations recommended by their authors. We did not consider GAN-based models such as PassGAN~\cite{passgan} because past research suggested that they need more guesses to obtain the same performance as FLA. Moreover, we did not include some recently proposed methods~\cite{chunk_level,newRules}, due to the lack of source code released.

\subsubsection{Comparing PSMs} 

One of the main conceptual forms for password strength evaluation is to simulate adversarial password guessing~\cite{backoff,minauto}. We compare the performance of the lightweight PassTSL meter with other strength meters described as follows.

\textbf{FLA-based PSM.} Authors of~\cite{fla} used a quantized and compressed version of FLA for their PSM, which can lead to worse performance. In our experiments, we used the uncompressed version of FLA so that the FLA-based PSM will demonstrate its best performance.

\textbf{zxcvbn.} zxcvbn~\cite{zxcvbn} is considered one of the best PSMs that rely on manual rules, statistical methods, and plain-text dictionaries.

The Yahoo!\ PSM used in~\cite{fla} and CKL\_PCFG PSM used in~\cite{chunk_level} were excluded because the former was far inferior to zxcvbn and the latter was only partially open-sourced. The ground truth is provided via the MinGuess method~\cite{minauto}. It is an idealized approach and represents the most conservative security results. A password is considered cracked as long as it is guessed by any of the guessing approaches.

\subsubsection{Evaluation Results on Password Guessing}
\label{section:pretraining-3-3}

Figure~\ref{fig:pretrain-2} shows the performances of PassTSL and other five SOTA guessing models when attacking 17173, 178, Tianya, MyHeritage, RockYou and Twitter. We can draw the following conclusions.

\textbf{PassTSL outperforms other five guessing models when attacking Chinese password databases.} As presented in Figures~\ref{fig:pretrain-2}(a)-(c), the simulation curves of $\text{PassTSL}_{\text{Small}}^{\text{COMB\_100M}}$ are at the top, indicating that PassTSL performs the best when attacking Chinese passwords.

We took 1,000 points uniformly from each curve in logarithmic coordinates and calculated the average of their differences to measure the performance more quantitatively. Specifically, when attacking 17173, the predicted coverage of PassTSL is \textbf{4.76\%} higher than FLA, \textbf{12.72\%} higher than 6-gram Markov, and \textbf{12.74\%} higher than backoff Markov at maximum around $10^{11}$ guesses, where \textbf{passwords almost surely will not survive credible offline attacks}~\cite{threshold_strength}. When the target database is 178, the predicted coverage of PassTSL is \textbf{5.86\%} higher than FLA, \textbf{15.46\%} higher than 6-gram Markov, and \textbf{15.87\%} higher than backoff Markov at the maximum point. When attacking Tianya, the predicted coverage rate of PassTSL is \textbf{4.11\%} higher than FLA, \textbf{12.29\%} higher than 6-gram Markov, and \textbf{10.84\%} higher than backoff Markov at the maximum point. PassTSL also outperforms PCFG$_{\text{Se}}$ with over \textbf{35\%} (\textbf{64.69\%} against Tianya on $10^{18}$ guesses) and PCFG$_{\text{Ori}}$ with over \textbf{25\%} for all Chinese databases.

\textbf{PassTSL is better when attacking English databases at larger guessing numbers.} Observing Figures~\ref{fig:pretrain-2}(d)-(f), the simulation curves of $\text{PassTSL}_{\text{Small}}^{\text{COMB\_100M}}$ are lower than the ones of PCFG$_{\text{Se}}$ and PCFG$_{\text{Ori}}$ when the guessing number is less than $10^8$. However, they remain the highest over $10^{10}$ when the target database is MyHeritage and over $10^{9}$ guesses for other target databases, while PCFG curves look saturated. These results demonstrate that although PCFG-based methods perform best on small-scale guesses when attacking English passwords, the sparsity defect of PCFG is gradually exposed as the guessing number increases. Some templates or patterns of target passwords were not learned by PCFG from the training data. However, PassTSL will significantly exceed PCFG since it builds a more complicated password model than the simpler PCFG-based one. Another network-based method FLA is inferior to PassTSL in all scenarios, reflecting the advantage of self-attention operations over the circular mechanism for password modeling.

\subsubsection{Evaluation Results on PSMs}

We quantized and compressed $\text{PassTSL}_{\text{Small}}^{\text{COMB\_100M}}$ to make it closer to a real-world PSM that can run from a web browser. The compressed PSM represents a more conservative performance of PassTSL-based PSMs. Table~\ref{tab:compressed-model} shows the size of our PassTSL PSM after quantization and lossless compression.

\begin{table*}[!htb]
\centering
\caption{The compressed model sizes and the loading times.}
\label{tab:compressed-model}
\begin{threeparttable}
    \begin{tabular}{c c c c c}
        \toprule
        Name & Compress & Model Size & Client Size$^a$ & Loading Time (seconds)\\
        \midrule
        $\text{PassTSL}_{\text{Small}}$ & - & 18.4 MB & 37.8MB & 1.86\\
        $\text{PassTSL}_{\text{Small}}^{\text{fp16}}$ & fp16 & 9.26 MB & 28.2MB & 1.55\\
        $\text{PassTSL}_{\text{Small}}^{\text{fp16z}}$ & fp16+zip & 8.48 MB & 27.6MB & 2.34\\
        $\text{PassTSL}_{\text{Small}}^{\text{int8}}$ & int8 & 4.75 MB & 23.5MB & 1.12\\
        $\text{PassTSL}_{\text{Small}}^{\text{int8z}}$ & int8+zip & 3.97 MB & 22.9MB & 1.49\\
        \bottomrule
    \end{tabular}
    \begin{tablenotes} 
        \footnotesize
        \item[a] The client size refers to the transferred data size. The loading time refers to the time when Google Chrome was used.
    \end{tablenotes}
\end{threeparttable}
\end{table*}

\begin{figure*}[htb]
\centering
\begin{subfigure}[b]{0.3\linewidth}
    \centering
    \includegraphics[width=\linewidth]{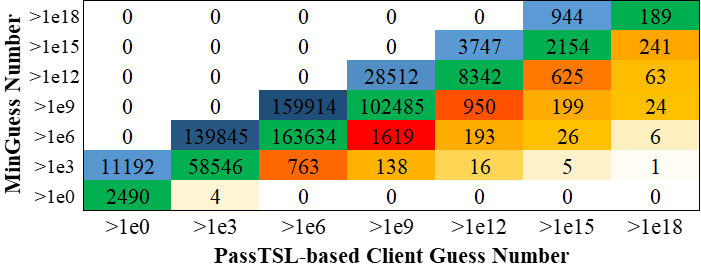}
    \caption{PassTSL}
\end{subfigure}
\hfill
\begin{subfigure}[b]{0.3\linewidth}
    \centering
    \includegraphics[width=\linewidth]{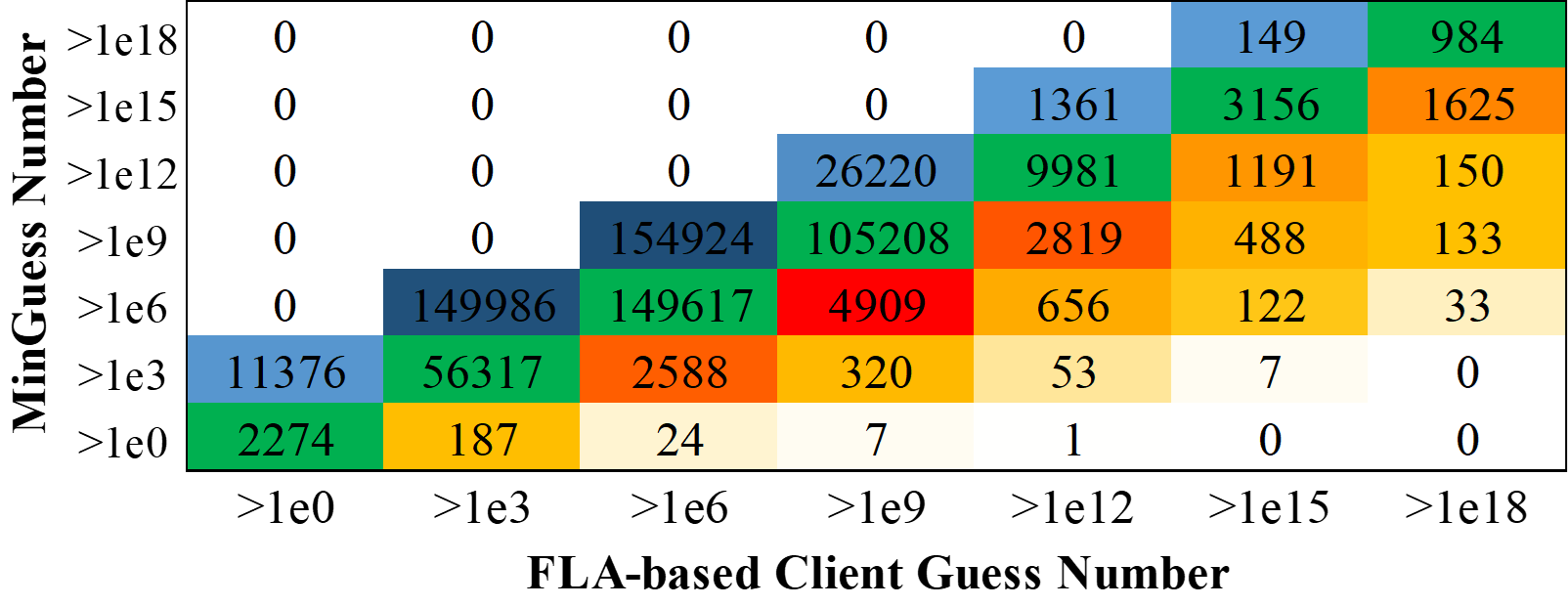}
    \caption{FLA}
\end{subfigure}
\hfill
\begin{subfigure}[b]{0.3\linewidth}
    \centering
    \includegraphics[width=\linewidth]{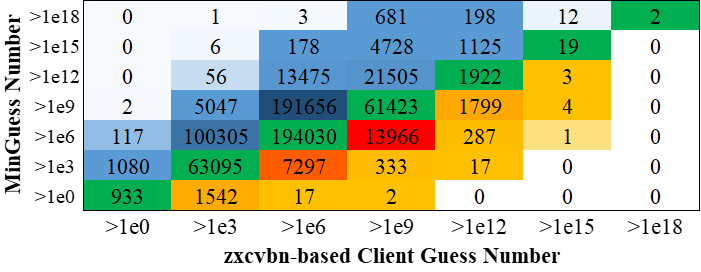}
    \caption{zxcvbn}
\end{subfigure}
\caption{Guessing numbers of the three tested PSMs compared against MinGuess. Red cells are important because they indicate how many passwords are overestimated by a PSM (denoted as unsafe errors). Underestimations of strength are shown in blue cells (called safe errors) and accurate estimations are shown in green cells. The chromatic intensity rises with the number of passwords.}
\label{fig:client-2}
\end{figure*}

We compared the accuracy of our lightest PSM, PassTSL-Small-int8z, to other PSMs in Figure~\ref{fig:client-2}. Here, we scale the output of PassTSL and FLA down to ensure that they made as many safe errors (under-estimated password strength) as zxcvbn (in this case, both PassTSL- and FLA-based PSMs gave more conservative results). The factors were 42 and 68, respectively.

From Figure~\ref{fig:client-2}, PassTSL-Small-int8z is more accurate than the other two PSMs: it has the lowest unsafe errors (over-estimated password strength), while safe errors are aligned with others. It makes significantly fewer (almost half for all red cells) unsafe errors than FLA. Moreover, although zxcvbn is more accurate when evaluating strong passwords ($>10^{13}$), its unsafe errors often appear at lower and mid-ranged guessing numbers, which could mislead users to choose weaker passwords more often.

\section{PassTSL: Finetuning}
\label{section:finetuning}

As explained in the previous section, our PassTSL base-line model pretrained using the COMB has already demonstrated superior performance over some SOTA methods. In this section, we show that even better performance can be achieved by finetuning the pretrained model. We explore methods for finetuning PassTSL and report the impact of various finetuning strategies.

\subsection{The Effects of Database Properties}
\label{section:finetuning-1}

In general, attackers know the language most users speak and the service type of a target website. We were interested in knowing if these two properties can positively affect the finetuning results. Wang et al.~\cite{birthday} showed that the native language plays an important role in users' password composition behaviors and users speaking different languages have noticeable different password structural patterns. For instance, English-speaking users prefer letters more than Chinese users, but less so on digits. We also noticed that, apart from listing the service types of websites from which the password databases were leaked, authors of past studies~\cite{personal_info,birthday,chunk_level} did not discuss the impact of website service type in more details. Our work reported here fills this gap.

Together with the findings in Section~\ref{section:pretraining-settings}, we consider the attack on a specific password database as a downstream task on $\text{PassTSL}_{\text{Small}}^{\text{COMB\_100M}}$. We expect that the password cracker could enhance password modeling performance with small finetuning costs while preserving common password knowledge learned in the pretraining phase. Intuitively, passwords taken from a website sharing the same language and service type properties as the target website are ideal sources for finetuning. We designed three scenarios to investigate the effects of both properties, where one million passwords were randomly selected from a finetuning database to attack a target database.

\textbf{The same service type (social media), but different languages (Chinese to English) (Scenario A).} $\text{PassTSL}_{\text{Small}}^{\text{COMB\_100M}}$ is finetuned by Tianya\_1M to attack RockYou and Twitter.

\textbf{The same language (English), but different service types (Scenario B).} $\text{PassTSL}_{\text{Small}}^{\text{COMB\_100M}}$ is finetuned by Twitter\_1M to attack MyHeritage (social media to information service) and by Tianya\_1M to attack 178 (social media to gaming website).

\textbf{The same language (English) and the same service type (social media) (Scenario C).} $\text{PassTSL}_{\text{Small}}^{\text{COMB\_100M}}$ is finetuned by 17173\_1M to attack 178 (Chinese gaming websites) and by Twitter\_1M to attack RockYou.

\begin{figure*}[htb]
\centering
\begin{subfigure}[b]{0.3\linewidth}
    \centering
    \includegraphics[width=\linewidth]{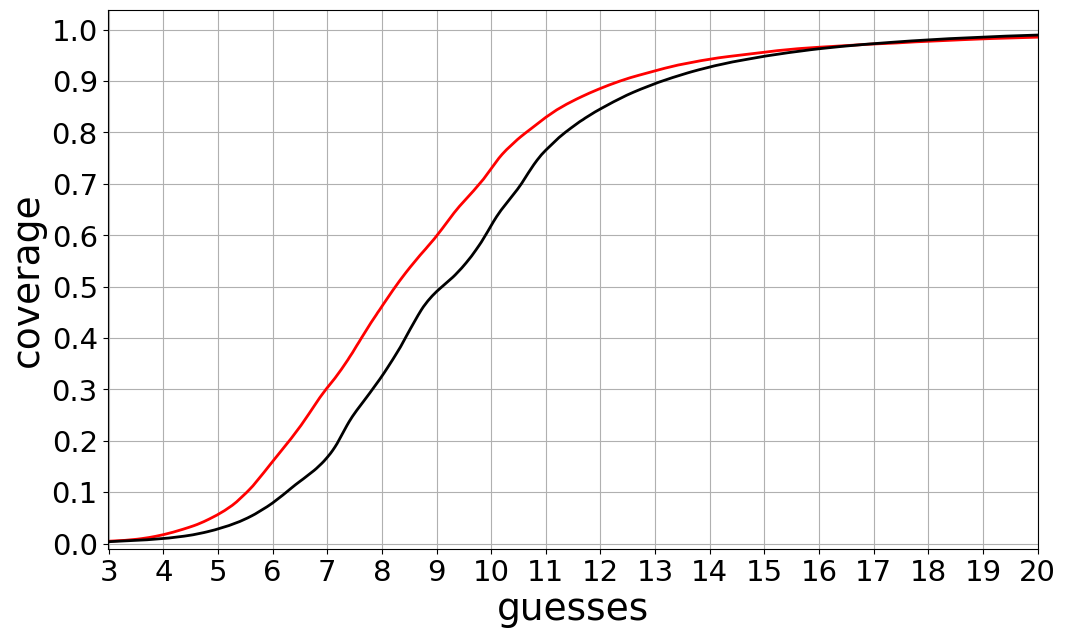}
    \includegraphics[width=\linewidth]{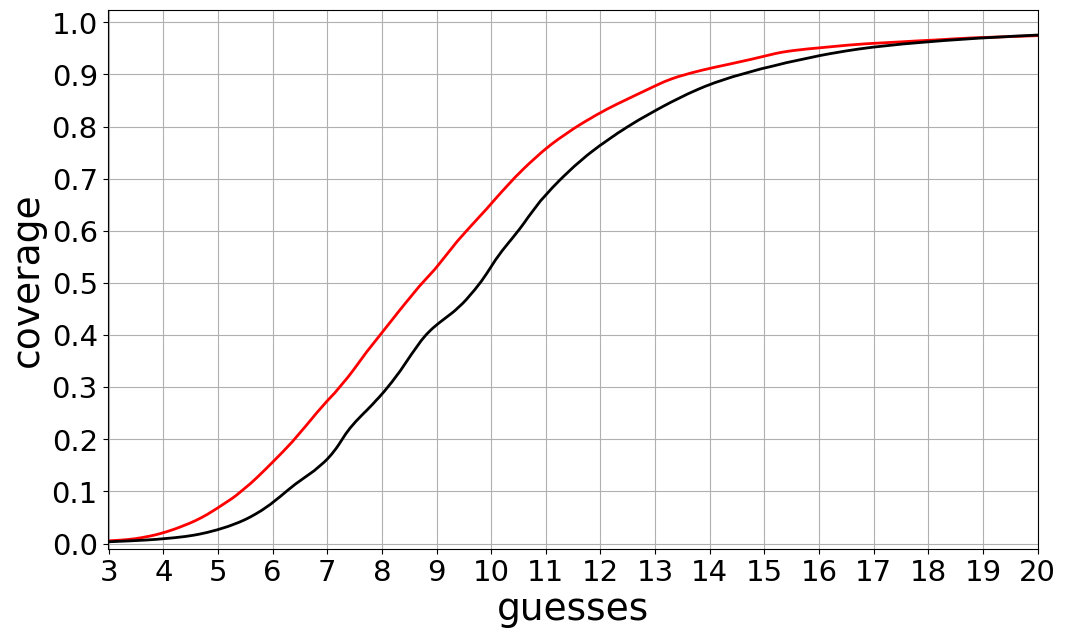}
    \caption{Scenario A}
\end{subfigure}
\hfill
\begin{subfigure}[b]{0.3\linewidth}
    \centering
    \includegraphics[width=\linewidth]{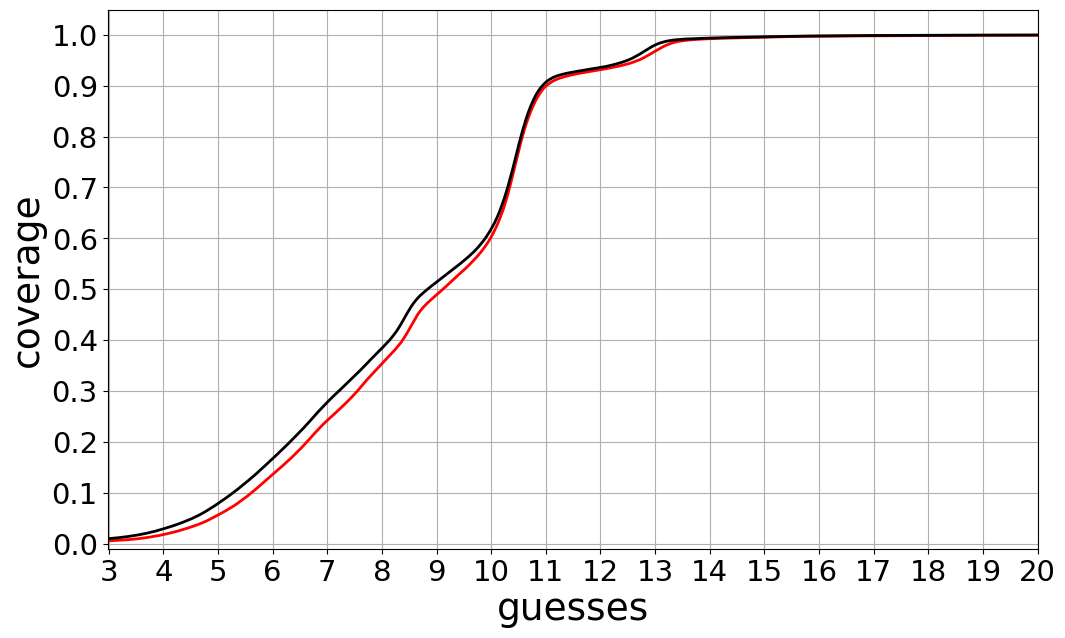}
    \includegraphics[width=\linewidth]{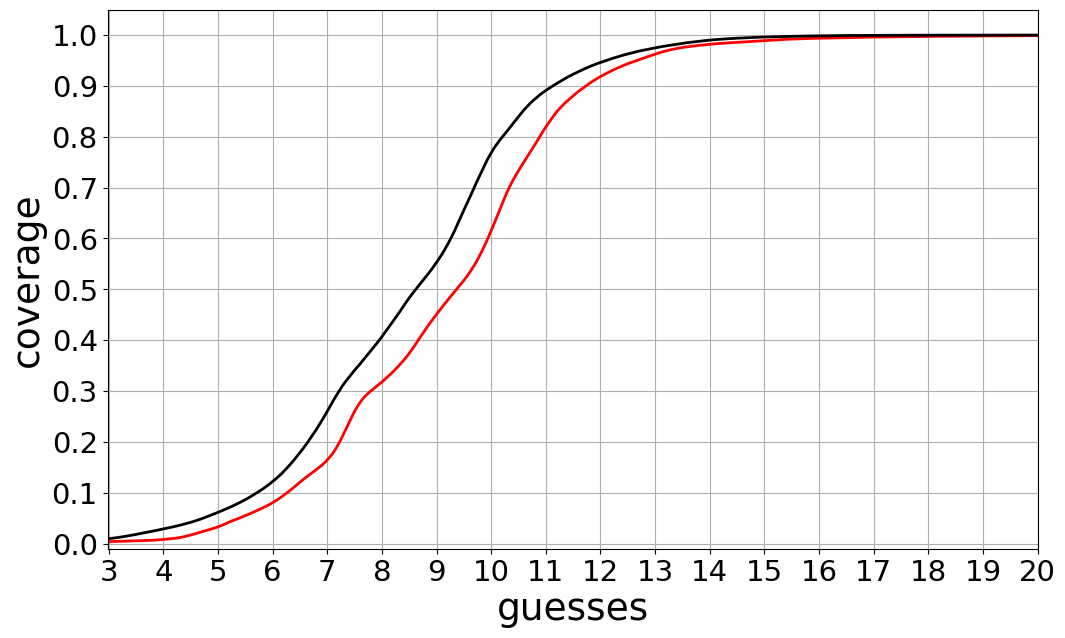}
    \caption{Scenario B}
\end{subfigure}
\hfill
\begin{subfigure}[b]{0.3\linewidth}
    \centering
    \includegraphics[width=\linewidth]{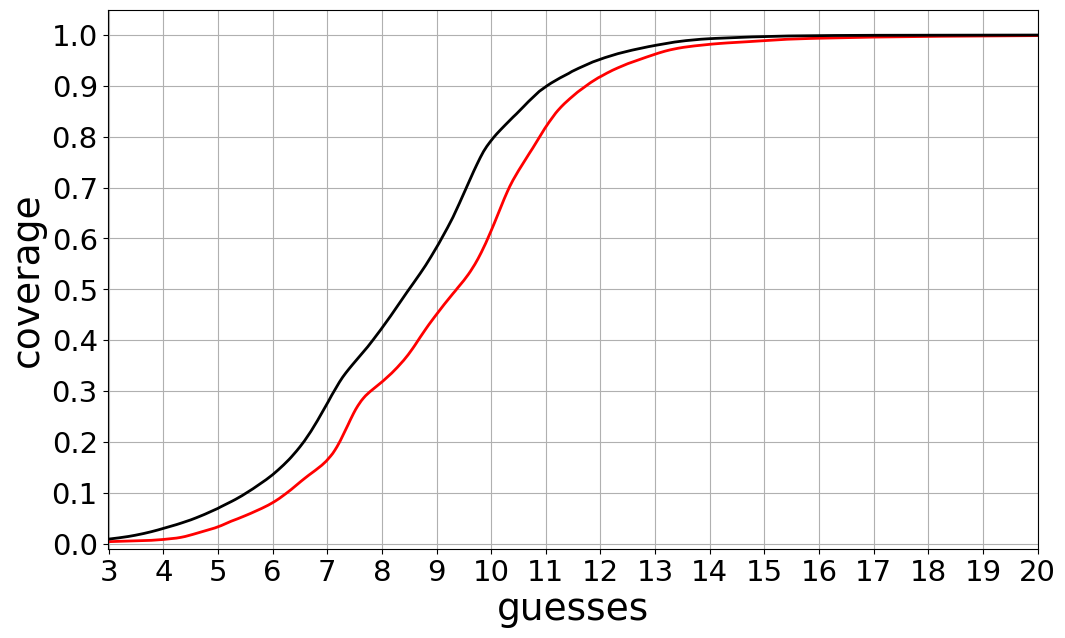}
    \includegraphics[width=\linewidth]{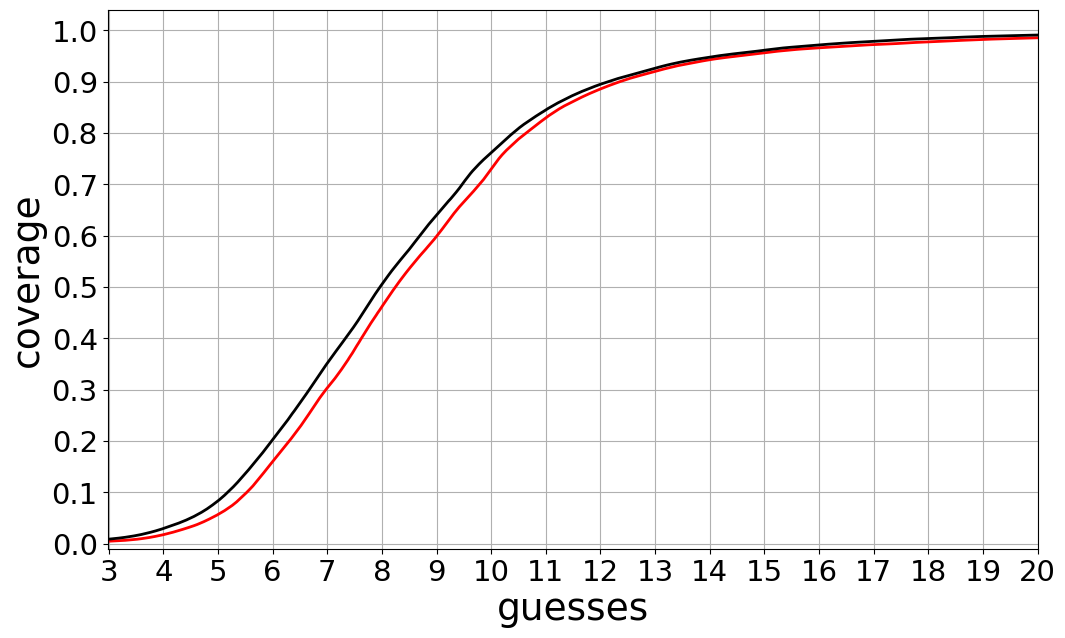}
    \caption{Scenario C}
\end{subfigure}
\caption{Performances of the finetuned PassTSL instances in the three scenarios for investigating the effects of the database attributes: \drawlegend{black} $\text{PassTSL}_{\text{Small}}^{\text{FT}}$, against \drawlegend{red} $\text{PassTSL}_{\text{Small}}^{\text{COMB\_100M}}$.}
\label{fig:finetune-1}
\end{figure*}

As shown in Figure~\ref{fig:finetune-1}, the curves in sub-figure (a) show that it will lead to a decrease on performance if finetuning only based on the service type of the target website, while results in sub-figures (b-c) prove that \textbf{the user language has a positive effect in the finetuning stage}: the coverage rate of $\text{PassTSL}_{\text{Small}}^{\text{FT}}$ (FT = finetuning) is on average 3.33\% higher than that of $\text{PassTSL}_{\text{Small}}^{\text{COMB\_100M}}$ in sub-figure (b), and 4.07\% in sub-figure (c).

\subsection{The Size of Finetuning Passwords}
\label{section:finetuning-2}

The results in the previous subsection are based on one million finetuning passwords. This amount of data may still considered too high, so we also investigated if a smaller amount of finetuning data can achieve a sufficiently good performance.

\begin{figure*}[th]
\centering
\begin{subfigure}[b]{0.45\linewidth}
    \centering
    \includegraphics[width=\linewidth]{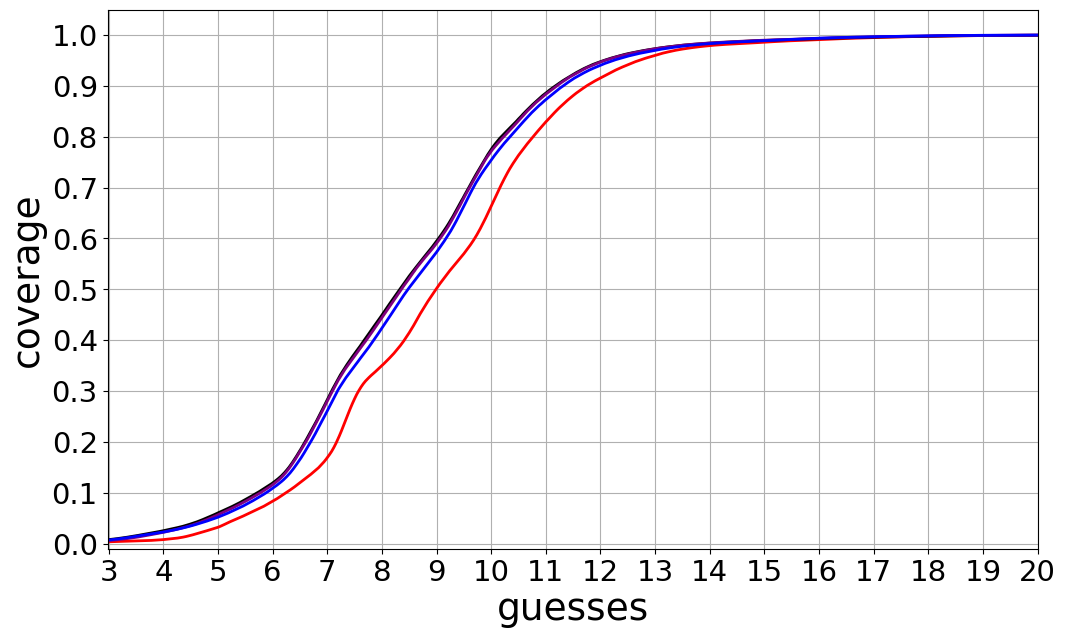}
    \caption{COMB-17173}
\end{subfigure}
\hfill
\begin{subfigure}[b]{0.45\linewidth}
    \centering
    \includegraphics[width=\linewidth]{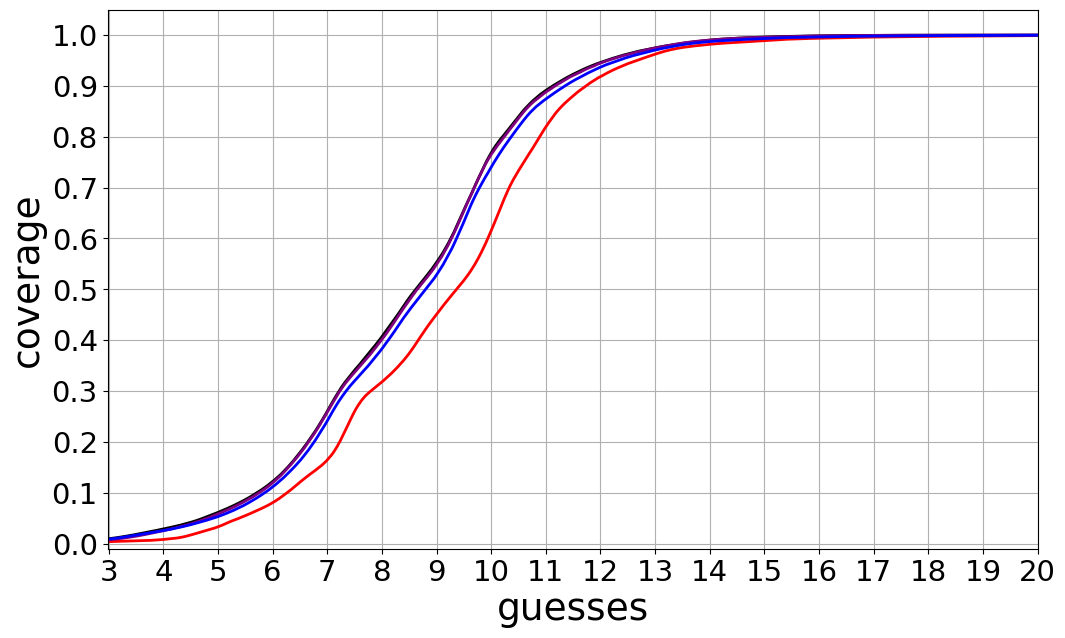}
    \caption{COMB-178}
\end{subfigure}
\caption{Performances of two PassTSL instances finetuned by a smaller database, \drawlegend{violet} $\text{PassTSL}_{\text{Small}}^{\text{FT\_100K}}$ and \drawlegend{blue} $\text{PassTSL}_{\text{Small}}^{\text{FT\_10K}}$, compared against two base-line instances, \drawlegend{red} $\text{PassTSL}_{\text{Small}}^{\text{COMB\_100M}}$ and \drawlegend{black} $\text{PassTSL}_{\text{Small}}^{\text{FT\_1M}}$.}
\label{fig:finetune-3}
\end{figure*}

Using the Tianya database as an example, 10K and 100K passwords were randomly selected, satisfying the requirements in Section~\ref{section:pretraining-settings-experimental}, denoted as Tianya\_10K and Tianya\_100K. They were new password sets to finetune $\text{PassTSL}_{\text{Small}}^{\text{COMB\_100M}}$, respectively, while target databases were 17173 and 178. The finetuned models, denoted by $\text{PassTSL}_{\text{Small}}^{\text{FT\_10K}}$ and $\text{PassTSL}_{\text{Small}}^{\text{FT\_100K}}$, were compared with $\text{PassTSL}_{\text{Small}}^{\text{FT\_1M}}$ and $\text{PassTSL}_{\text{Small}}^{\text{COMB\_100M}}$.

Experimental results are provided in Figure~\ref{fig:finetune-3}. For both target databases, curves of $\text{PassTSL}_{\text{Small}}^{\text{FT\_1M}}$ and $\text{PassTSL}_{\text{Small}}^{\text{FT\_100K}}$ are approximately identical, while visible gaps exist for $\text{PassTSL}_{\text{Small}}^{\text{FT\_10K}}$. In particular, $\text{PassTSL}_{\text{Small}}^{\text{FT\_100K}}$ is on average 0.13\% lower than $\text{PassTSL}_{\text{Small}}^{\text{FT\_1M}}$ on 17173 and 0.14\% lower on 178, while $\text{PassTSL}_{\text{Small}}^{\text{FT\_10K}}$ is 0.77\% and 0.91\% lower. However, coverage rates of $\text{PassTSL}_{\text{Small}}^{\text{FT\_100K}}$ is still 3.4\% and 3.7\% higher than $\text{PassTSL}_{\text{Small}}^{\text{COMB\_100M}}$, respectively. \textbf{The results show that, for the finetuning stage, only 0.1\% of the pretraining data size can be sufficient to obtain a good level of performance improvement.} The finding echoes the observation in Section~\ref{section:pretraining-settings} that using COMB for pretraining can lead to a pretrained model biased towards English passwords.

From the experiments in Sections~\ref{section:finetuning-1} and \ref{section:finetuning-2}, we conclude that:
\begin{itemize}
\item The finetuning process can enhance PassTSL's ability to guess passwords.

\item The user language of the target website can be used to achieve better finetuning results, if the finetuning data share the same language as the target website. Considering that this property of a website is mostly public knowledge, an immediate advice to users is that they should try to diversify ways to define their passwords, particularly to avoid using elements that are more typical in the dominating language of a website.

\item To achieve a good performance, the finetuning stage just needs as little as 0.1\% of the amount of pretrained passwords.
\end{itemize}

\subsection{How to Select Finetuning Password Database}

\begin{table}[htb]
\centering
\caption{JS divergence and finetuning settings}
\label{tab:finetune-dataset}
\begin{threeparttable}
\begin{tabular}{c c c c c c c}
\toprule
Database & Usage & 17173 & 178 & MyHeritage & RockYou & Twitter\\
\midrule
COMB\_100M & Pretraining & 0.2288 & 0.256 & 0.1339 & 0.0645 & 0.0206\\
\midrule
Tianya\_1M & Finetuning & 0.0268 & 0.0511 & - & - & -\\
Twitter\_1M & Finetuning & -$^a$ & - & 0.1311 & 0.057 & -\\
RockYou\_1M & Finetuning & - & - & 0.1431 & - & 0.057\\
\bottomrule
\end{tabular}
\begin{tablenotes} 
\item[a] `-' means we did not use this pair of databases for our experiments.
\end{tablenotes}
\end{threeparttable}
\end{table}

Based on the experimental results in Section~\ref{section:pretraining}, it is recommended that COMB be used as the pretraining database. Heuristically, we suggest selecting the finetuning database so that it is more similar to the target database than the pretraining database is, so that the finetuning process can add more specific information about target passwords to the finetuned PassTSL model. To measure the similarity between two password databases, we propose to use the JS (Jensen-Shannon) divergence~\cite{JSD} calculated based on the union of all 3-grams in the two password databases.

To validate the above heuristic idea, some experiments were conducted based on JS divergence values between selected pairs of password databases given in Table~\ref{tab:finetune-dataset}. The finetuned PassTSL in each experiment is denoted as $\text{PassTSL}_{\text{Small}}^{\text{FT}}$.

\begin{figure*}[htb]
\centering
\begin{subfigure}[t]{0.3\linewidth}
    \centering
    \includegraphics[width=\linewidth]{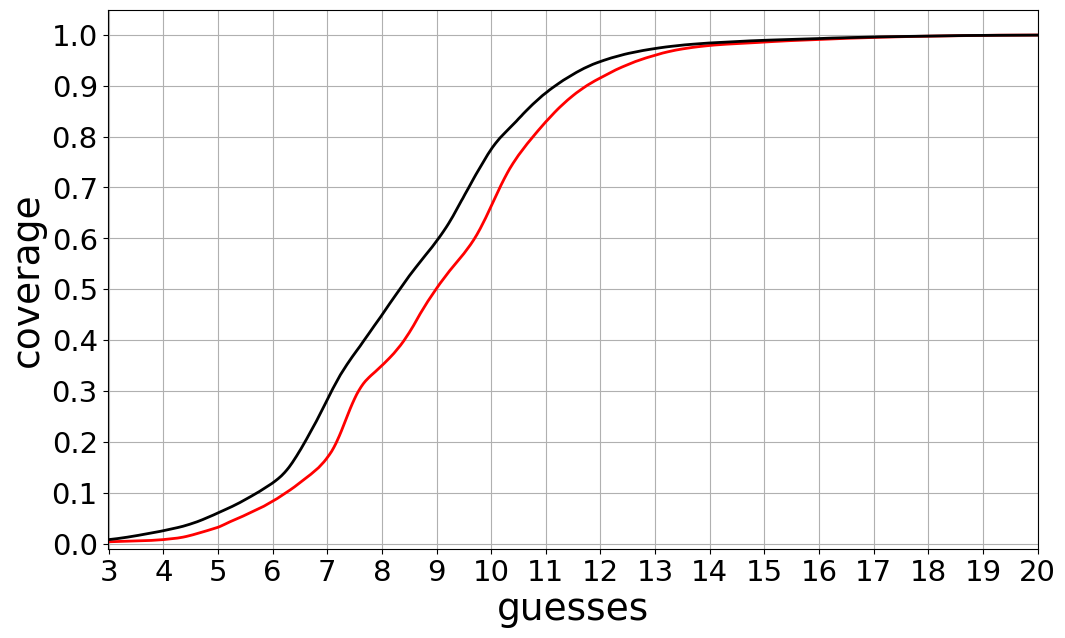}
    \includegraphics[width=\linewidth]{part2/2-1/Tianya/Finetune-CN-178.png}
    \caption{Tianya-finetuned PassTSL attacking 17173 (top) \& 178 (bottom)}
\end{subfigure}
\hfill
\begin{subfigure}[t]{0.3\linewidth}
    \centering
    \includegraphics[width=\linewidth]{part2/2-1/Twitter/Finetune-EN-Myheritage.png}
    \includegraphics[width=\linewidth]{part2/2-1/Twitter/Finetune-EN-Rockyou.png}
    \caption{Twitter-finetuned PassTSL attacking MyHeritage (top) \& RockYou (bottom)}
\end{subfigure}
\hfill
\begin{subfigure}[t]{0.3\linewidth}
    \centering
    \includegraphics[width=\linewidth]{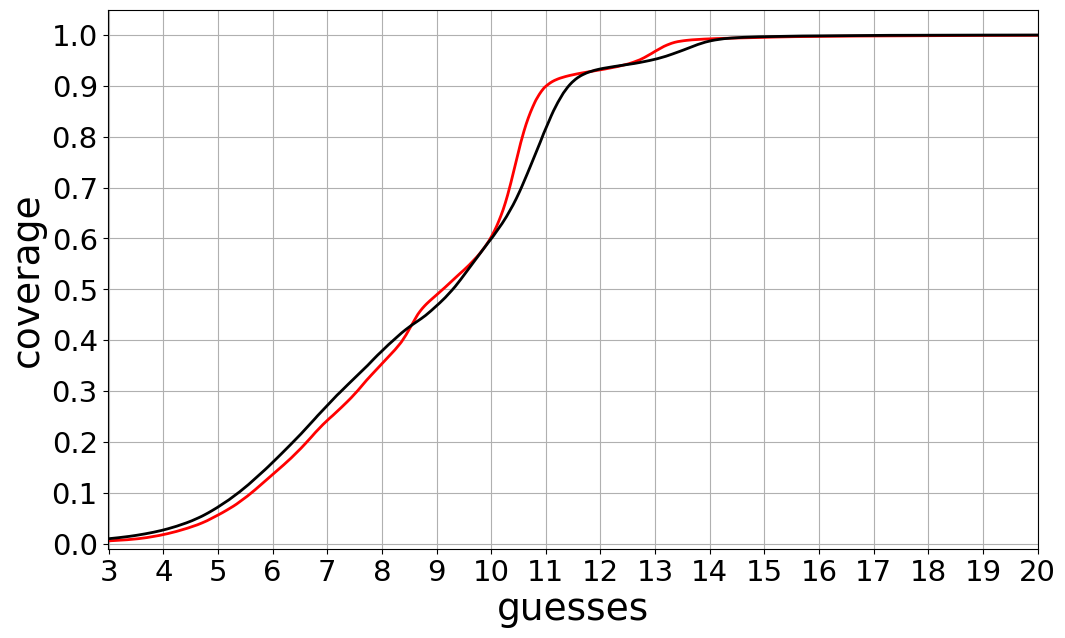}
    \includegraphics[width=\linewidth]{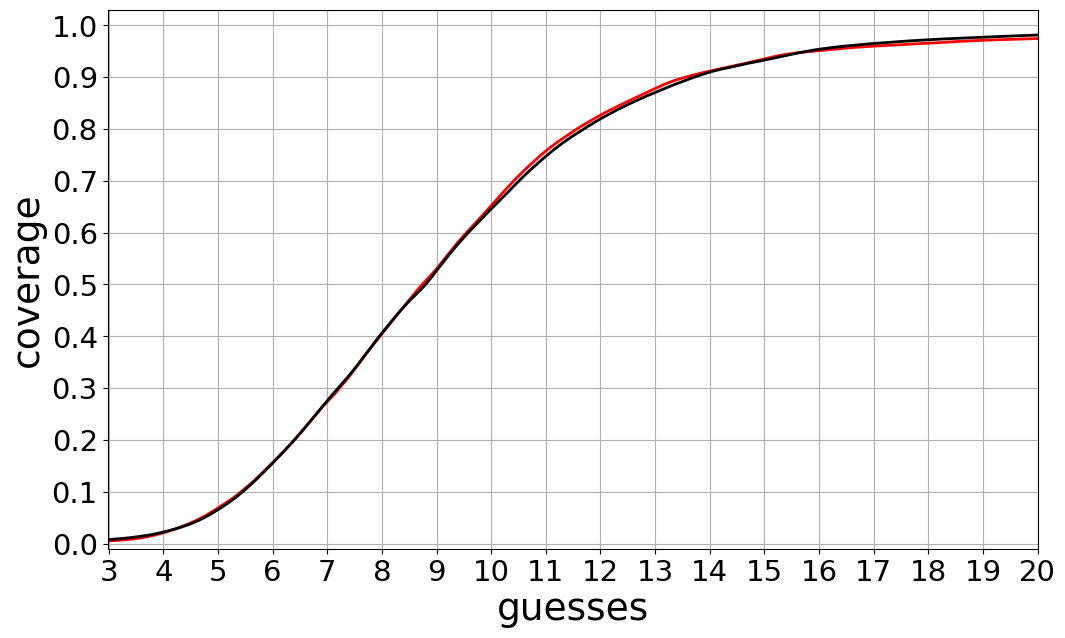}
    \caption{RockYou-finetuned PassTSL attacking MyHeritage (top) \& Twitter (bottom)}
\end{subfigure}
\caption{Performances of finetuned PassTSL instances compared against those without finetuning: \drawlegend{black} $\text{PassTSL}_{\text{Small}}^{\text{FT}}$ and \drawlegend{red} $\text{PassTSL}_{\text{Small}}^{\text{COMB\_100M}}$.}
\label{fig:finetune-2}
\end{figure*}

Figure~\ref{fig:finetune-2} presents experimental results. As shown in in Sub-figure (a), compared with the pretrained model $\text{PassTSL}_{\text{Small}}^{\text{COMB\_100M}}$, the model $\text{PassTSL}_{\text{Small}}^{\text{FT}}$ finetuned using the Tianya database performs significantly and consistently better when attacking 17173 and 178 (the coverage rate increases by 12.13\% on $10^{9}$ guesses on 17173 and by 15.77\% on $10^{9}$ guesses on 178). Such results can be predicted from the JS divergence values: $\text{JS}(\text{Tianya}, \text{17173})<\text{JS}(\text{COMB\_100M}, \text{17173})$ and $\text{JS}(\text{Tianya}, \text{178})<\text{JS}(\text{COMB\_100M}, \text{178})$. When the Twitter database was used for finetuning, $\text{PassTSL}_{\text{Small}}^{\text{FT}}$ also beats $\text{PassTSL}_{\text{Small}}^{\text{COMB\_100M}}$ on MyHeritage and RockYou as shown in Sub-figure (b). The coverage rate increases by 3.9\% on $10^8$ guesses and 4.96\% on $10^7$ guesses. Such results could also be predicted from the JS divergence values: $\text{JS}(\text{Twitter}, \text{MyHeritage})<\text{JS}(\text{COMB\_100M}, \text{MyHeritage})$ and $\text{JS}(\text{Twitter}, \text{RockYou})<\text{JS}(\text{COMB\_100M}, \text{RockYou})$. The lower improvement may be explained by the fact that the JS divergence values between the Twitter database and the two target databases are just slightly smaller than the values between COMB\_100M and the two target databases. When the RockYou database was used for finetuning, the performance of the finetuned model $\text{PassTSL}_{\text{Small}}^{\text{FT}}$ behaves very similar to $\text{PassTSL}_{\text{Small}}^{\text{COMB\_100M}}$, which can be explained by the fact that the RockYou database does not have a smaller JS divergence value with the target databases than COMB\_100M, as shown in Table~\ref{tab:finetune-dataset}, therefore the contribution of the finetuning becomes more marginal (if any).

In summary, the results in Figure~\ref{fig:finetune-2} largely validate the effectiveness of using the JS divergence as a quantitative metric to guide the selection of the finetuning database.

\section{Conclusion}

This paper presents PassTSL, a deep learning model for password modeling and guessing, based on the pretraining-finetuning two-staged learning framework. Our model aims to extract universal password features through a self-attention mechanism. Extensive experiments have proved that PassTSL is superior to other SOTA password modeling and cracking methods and is also practical to support a password strength meter. It is also shown that the finetuning phase can help improve the performance even further if the finetuning database is properly selected, and the amount of finetuning data needed is very light (e.g., 0.1\% of the pretraining data).

\bibliographystyle{splncs04}
\bibliography{paper}

\end{document}